\documentclass[11pt,a4paper]{article}
\usepackage[T1]{fontenc}
\usepackage{iftex}
\ifPDFTeX\usepackage[utf8]{inputenc}\fi
\usepackage{lmodern}
\usepackage[a4paper,margin=25mm]{geometry}
\usepackage{microtype}
\usepackage{amsmath,amssymb,mathtools}
\usepackage{graphicx}
\usepackage{booktabs,array,tabularx,longtable,float}
\usepackage[font=small,labelfont=bf]{caption}
\usepackage[numbers,sort&compress]{natbib}
\usepackage{xurl}
\usepackage{xcolor}
\usepackage[unicode,colorlinks=true,linkcolor=blue!45!black,citecolor=blue!45!black,urlcolor=blue!45!black]{hyperref}
\usepackage{authblk}
\usepackage{orcidlink}
\usepackage{hyperref} 

\hypersetup{
  pdftitle={Directional memory of early spectral selection during spinodal decomposition in finite systems},
  pdfauthor={Boliang Yu, Second Author, Third Author}
}

\title{Directional memory of early spectral selection during spinodal decomposition in finite systems}

\author[1]{Boliang Yu\,\orcidlink{0009-0006-4664-5068}\thanks{
Corresponding author: \href{mailto:yuboliang@sjtu.edu.cn}{yuboliang@sjtu.edu.cn}}}
\author[1]{Ruixin Zhou\,\orcidlink{0009-0004-9303-6674}}
\author[2]{Zheng Zhang\,\orcidlink{0000-0002-4152-8552}}

\affil[1]{School of Physics and Astronomy, Shanghai Jiao Tong University, Shanghai 200240, China}
\affil[2]{Department of Physics, School of Science, Lanzhou University of Technology, Lanzhou 730050, China}

\date{}
\begin{document}
\maketitle
\begin{abstract}
Cahn--Hilliard phase separation turns small composition fluctuations into coarsening domains. In Fourier space, the pattern forms a ring whose angular intensity favors an axis. We ask whether the axis favored early in an evolution remains related to the axis favored later in the same evolution. We analyze two-dimensional simulations with isotropic initial fluctuations and no imposed direction. We introduce $Q_2$, a normalized angular average over the Fourier spectrum, as a measure of the strength and direction of spectral anisotropy. Linear growth amplifies different wave numbers at different rates, selecting an early axis from the angular imbalances in the initial spectrum. This axis remains correlated with the late axis during coarsening. At a fixed time, shallow quenches show stronger early--late alignment. This advantage disappears when the quenches are compared at similar stages of domain growth. Adding the early $Q_2$ to a prediction based on quench depth, mean composition, and box size reduces the mean absolute error for late $Q_2$ by $3.53\%$. This reduction disappears when only the magnitude of the early $Q_2$ is retained, or when its direction is replaced by the direction from another evolution. The early Fourier intensity therefore carries information about the later state of the same evolving field.
\end{abstract}

\section{Introduction}\label{sec:introduction}

Many physical patterns begin in the same simple way: a nearly uniform material becomes unstable, small fluctuations grow, and a visible structure appears. When the governing laws are rotationally symmetric, they can select a characteristic length without imposing a preferred direction. A finite sample does not reproduce the ensemble average exactly. Its initial fluctuations are not equal in every direction, so one axis can become more prominent than the others \cite{Cahn1965,ShinozakiOono1993}. What happens to this finite-sample direction as the pattern becomes nonlinear is part of the broader problem of how initial fluctuations shape later structure.

Spinodal decomposition provides a direct setting for this question. After a homogeneous binary mixture is quenched into an unstable state, composition variations in a band of wave numbers grow, interfaces form, and diffusion drives the domains to coarsen \cite{CahnHilliard1958,Cahn1961,Cook1970,Joly1996}. The Cahn--Hilliard equation describes this sequence while conserving the mean composition, and its linearized form identifies the initially unstable wavelengths \cite{HohenbergHalperin1977}. At later times, phase-ordering theories and simulations describe the growth of a characteristic domain scale and the accompanying motion of spectral power toward smaller wave numbers \cite{Huse1986,Bray1994,Konig2021}. A characteristic length and a radially averaged structure factor therefore give a natural description of the scale evolution shared across realizations. By construction, however, the radial average discards the angular detail by which finite realizations can differ.

The full Fourier intensity retains that angular detail. Coherent-scattering studies have shown that a single evolving configuration produces a structured intensity pattern whose fluctuations can be followed in time \cite{Sutton1991,Malik1998,Brown1997,Brown1999}. Two-time correlations describe how configurations remain related during phase ordering \cite{Furukawa1990,YeungRaoDesai1996}, while a directional memory effect has been studied when uniaxial compression imposed a preferred direction on spinodal decomposition \cite{HashimotoIzumitani1996}. These studies establish that scattering can resolve temporal structure beyond the radial scale and motivate asking whether a much more compact angular feature remains identifiable.

A parallel computational literature asks how an earlier microstructure constrains its later evolution. Recent studies have used phase-field trajectories to predict later microstructures from earlier states \cite{Zapiain2021,Oommen2022,Rieger2024}. Their main task is to reproduce or forecast the evolving microstructure. In this paper, prediction tests whether an early angular measurement supplies information about the later state of the same evolution beyond what is known from the physical condition (quench depth, mean composition, and box size).

These two lines of work leave a specific gap. Existing correlation analyses do not isolate a single global axis selected by the initial angular imbalance of one finite evolution, and the cited phase-field prediction studies do not test whether such an axis adds information beyond the physical condition. We therefore ask whether an evolution under isotropic dynamics can retain a direction that arose from its own finite initial fluctuations. We then ask whether the early direction carries information about the later direction of the same evolution. Replacing the early direction with one from another evolution tests whether that information depends on the pairing.

\begin{figure}[H]
\centering
\includegraphics[width=\linewidth]{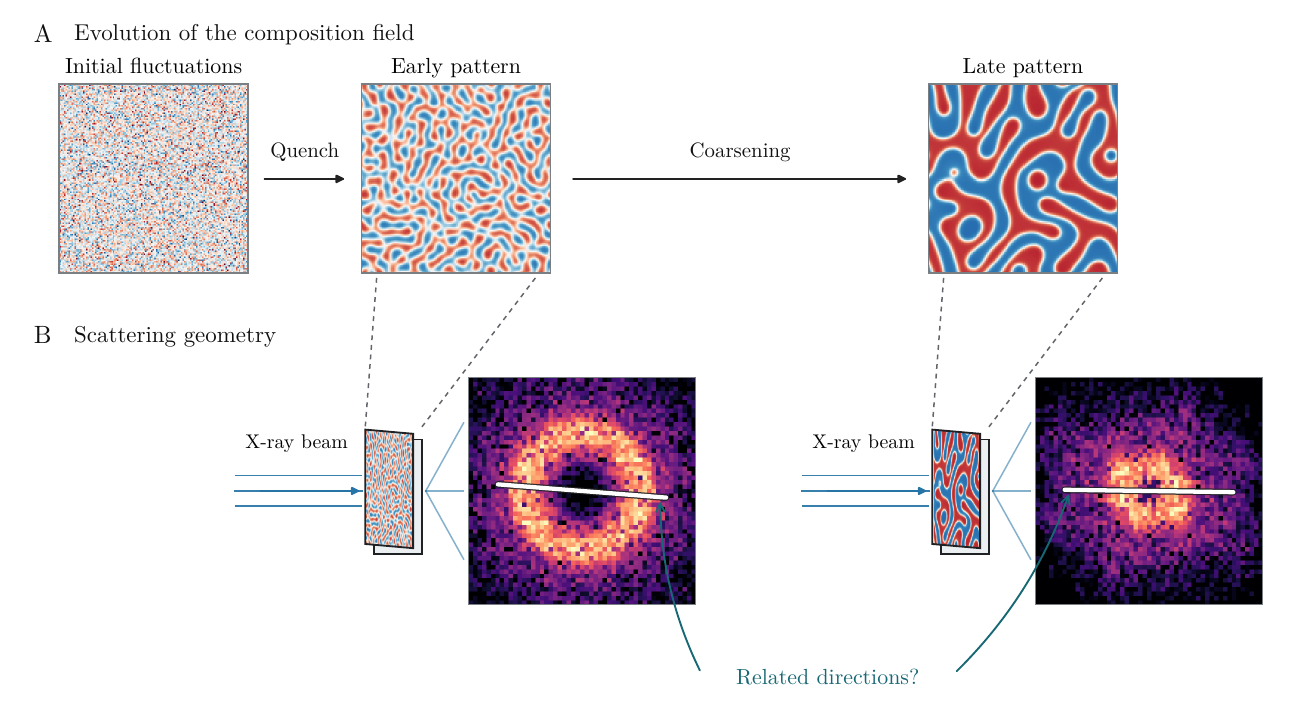}
\caption{\textbf{A scattering picture of the early--late direction question.} (A) Simulated composition fields at three stages of one Cahn--Hilliard evolution: initial fluctuations, an early domain pattern, and a late coarsened pattern. (B) Two separate schematic scattering geometries for the early and late fields. Each film face displays the corresponding field from (A), and each detector displays its Fourier intensity. White lines mark the spectral axes; the arrows identify the two axes being compared.}
\label{fig:overview}
\end{figure}

Figure~\ref{fig:overview} gives an intuitive scattering picture of the directional-memory question. A beam illuminates a phase-separating film, and a detector records the Fourier intensity of its composition pattern. As the domains grow, the intensity ring moves inward and its detailed pattern changes. The early and late rings can nevertheless be brightest along particular diameters. The analysis below asks whether those two diameters remain related within the same evolution. The film, beam, and detector illustrate the Fourier measurement; the fields themselves come from Cahn--Hilliard simulations.

We describe the spectral axis by $Q_2$, a normalized angular average over the Fourier spectrum:
\begin{equation*}
Q_2(t)=
\frac{\displaystyle\sum_{\mathbf k\in\mathcal K}
[S(\mathbf k,t)/|\mathbf k|]e^{2i\theta_{\mathbf k}}}
{\displaystyle\sum_{\mathbf k\in\mathcal K}
S(\mathbf k,t)/|\mathbf k|}
=|Q_2(t)|e^{2i\psi(t)}.
\end{equation*}
Here $S(\mathbf k,t)$ is the Fourier intensity, $\theta_{\mathbf k}$ is the polar angle of the wave vector, and $\mathcal K$ is the circular set of nonzero Fourier modes defined in Sec.~\ref{sec:observables}. The magnitude $|Q_2|$ measures how strongly the spectrum favors an axis, while $\psi=\tfrac12\arg Q_2$ gives the direction of that axis modulo $\pi$. Opposite Fourier directions therefore represent the same axis. Because the dynamics and the initial-noise distribution are rotationally symmetric, different evolutions select unrelated axes and the ensemble has no preferred direction. Directional memory can only appear as a relation between the early and late axes of the same evolution.

To test these questions, we simulate an ensemble of two-dimensional Cahn--Hilliard evolutions over several quench depths, mean compositions, and box sizes. We first follow the domain scale and spectral axis during coarsening. We compare early--late alignment across quenches at a fixed early time and at comparable stages of coarsening. We then propagate each actual initial spectrum with the linearized dynamics to determine how the early axis is selected and compare that calculation with the measured evolution. Finally, we ask whether the measured early $Q_2$ improves prediction of the late $Q_2$ beyond the information already supplied by the physical condition, and whether any improvement depends on using the direction from the same evolution.

The early and late axes remain related within the same evolution. Differences among quenches at a fixed time largely disappear when they are compared at similar stages of coarsening. Linear amplification of each actual initial spectrum explains how the early axis is selected. The angular error of this calculation explains why the calculated early axis aligns less closely with the late axis than the observed early axis does. After accounting for that error, the comparison detects no additional steering during nonlinear coarsening. The early direction also reduces the prediction error of the late $Q_2$ beyond a prediction based only on the physical condition, but only when it remains paired with the same evolution.

These results connect two descriptions that are usually considered separately. Rotational symmetry governs the ensemble, while a finite realization carries a directional history selected from its own initial fluctuations. Linear instability acts as the first selector of that history, and nonlinear coarsening transforms the morphology without completely removing it. The prediction comparison shows that this history belongs to an individual evolution. The results also show why comparisons across physical conditions should account for the stage of coarsening as well as elapsed time.

The remainder of the paper follows this argument. Section~\ref{sec:model-methods} introduces the Cahn--Hilliard model, defines the spectral measurements, and presents the design of each comparison. Section~\ref{sec:results} reports the scale growth, early--late alignment, linear calculation, and prediction results. Section~\ref{sec:discussion} develops their physical and methodological implications, and Section~\ref{sec:conclusion} states the conclusions. The appendices contain the numerical, statistical, and robustness details.

\section{Model and methods}\label{sec:model-methods}

\subsection{Cahn--Hilliard dynamics and simulation ensemble}\label{sec:model-setup}

To study whether a direction selected early in phase separation remains visible after coarsening, we evolve a conserved composition field with isotropic Cahn--Hilliard dynamics. Small random composition fluctuations are present initially. During the early linear instability, fluctuations in a band of wavelengths grow, and those near the fastest-growing wavelength are amplified most strongly. As the composition contrast becomes large, nonlinear terms limit this growth and interfaces form between the two phases. Diffusion then drives coarsening: small domains disappear, larger domains grow, and the dominant spectral power moves toward smaller wave numbers. At sufficiently late times in a finite square, the box size also limits the structures that can develop. These are successive parts of one evolution rather than sharply separated regimes.

The composition field $\phi(\mathbf x,t)$ evolves on a periodic square of side $L_{\mathrm{box}}$ according to the Cahn--Hilliard equation,

\begin{equation}\label{eq:1}
\begin{aligned}
F[\phi]&=\int\left[\frac r2\phi^2+\frac14\phi^4+\frac\kappa2|\nabla\phi|^2\right]d^2x,\\
\partial_t\phi&=M\nabla^2\mu,\qquad
\mu=r\phi+\phi^3-\kappa\nabla^2\phi.
\end{aligned}
\end{equation}

Here $r<0$ controls the quench depth, $\kappa$ sets the energetic cost of an interface, $M$ is the mobility, and $\mu$ is the chemical potential. The local free energy favors phase separation, while the gradient term opposes arbitrarily sharp interfaces. The Laplacian in the evolution equation describes diffusive transport and preserves the spatial mean $\bar\phi$ \cite{CahnHilliard1958,Cahn1961,HohenbergHalperin1977}. No directional field, flow, or stochastic forcing acts after initialization. The evolution decreases the free energy, $dF/dt=-M\int|\nabla\mu|^2d^2x$, while conserving the mean composition. Neither property introduces a preferred direction or requires the pattern to keep a fixed axis.

Linearization explains the early wavelength selection. Writing $\phi=\bar\phi+\delta\phi$ and retaining terms linear in the fluctuation $\delta\phi$ gives

\begin{equation}\label{eq:2}
\begin{aligned}
\partial_t\widehat{\delta\phi}_{\mathbf k}&=\sigma(k)\widehat{\delta\phi}_{\mathbf k},\\
\sigma(k)&=-Mk^2(a+\kappa k^2),\qquad a=r+3\bar\phi^2.
\end{aligned}
\end{equation}

The vector $\mathbf k$ is a Fourier wave vector and $k=|\mathbf k|$ is its magnitude. When $a<0$, modes with $0<k<\sqrt{-a/\kappa}$ grow, and the growth rate of the continuum equation is greatest at $k_*=\sqrt{-a/(2\kappa)}$ \cite{Cahn1961}. Because $\sigma$ depends on $k$ but not on the direction of $\mathbf k$, the linear instability favors wavelengths without imposing an axis. Once the nonlinear term becomes important, the Fourier modes no longer evolve independently: the growing composition contrast is limited, domains form, and those domains subsequently coarsen \cite{Konig2021}.

The numerical ensemble contains $1296=27\times48$ trajectories. A trajectory is the time evolution from one independently sampled initial field. We simulate $48$ trajectories for each combination of $r\in\{-0.5,-1,-1.5\}$, $N\in\{64,96,128\}$, and $\bar\phi\in\{0,0.15,0.30\}$; each triple $c=(r,N,\bar\phi)$ is called a condition. We use dimensionless units with $M=\kappa=\Delta x=1$, where $N$ is the number of grid points per side and $L_{\mathrm{box}}=N\Delta x$. The initial field is the prescribed mean plus independent Gaussian grid fluctuations of standard deviation $0.02$, then recentered to the prescribed mean.

The numerical update, saved times, and convergence checks are given in Appendices~\ref{app:A} and \ref{app:B}.

\subsection{Spectral measures of direction and length scale}\label{sec:observables}

To test directional persistence, we need a quantity that assigns both a strength and an axis to the Fourier spectrum of each stored field. We begin with the Fourier power spectrum, or structure factor,

\begin{equation}\label{eq:3}
S(\mathbf k,t)=\frac{1}{N^2}\left|\operatorname{DFT}[\phi(\mathbf x,t)-\bar\phi]_{\mathbf k}\right|^2.
\end{equation}

The value $S(\mathbf k,t)$ measures the power of the composition variation at wave vector $\mathbf k$, with the normalization specified in Eq.~\eqref{eq:3}. Let $\theta_{\mathbf k}$ be the angle of that wave vector. We evaluate the following quantities on the nonzero circular part $\mathcal K=\{\mathbf k:0<|\mathbf k|\leq\pi/\Delta x\}$ of the square Fourier grid. Within this range, $p_{\mathbf k}\propto S(\mathbf k,t)/|\mathbf k|$ is the normalized contribution of each Fourier mode, with $\sum_{\mathcal K}p_{\mathbf k}=1$. The circular cutoff excludes the corners of the square Fourier grid and sets the same upper wave-number bound in every direction. A wave-number shell contains modes within a narrow interval of $|\mathbf k|$. The factor $1/k$ approximately compensates for the increasing number of modes in shells of equal width; their contributions still depend on the spectral power. Appendix~\ref{app:A} gives the remaining numerical conventions. The two measurements needed for the main argument are

\begin{equation}\label{eq:4}
Q_2(t)=\sum_{\mathcal K}p_{\mathbf k}(t)e^{2i\theta_{\mathbf k}}
=A_S e^{2i\psi},
\qquad
L_1(t)=\frac{2\pi}{\sum_{\mathcal K}p_{\mathbf k}(t)|\mathbf k|}.
\end{equation}

The complex number $Q_2$ summarizes the twofold angular imbalance of the spectrum. Its magnitude $A_S=|Q_2|$ measures how strongly the power favors one axis, while $\psi=\arg(Q_2)/2$ gives that axis modulo $\pi$. The doubled angle makes opposite Fourier directions represent the same axis. For a striped pattern, this spectral axis is perpendicular to the stripes. The axis is constructed from Fourier intensity and is unrelated to the spatial Fourier phase. When $A_S$ is very small, the axis is poorly determined; this is why the principal alignment average below gives more weight to patterns with a clearer axis. The characteristic length $L_1$ summarizes the radial spectrum as a wavelength \cite{Bray1994}.

\subsection{Early--late comparison of the spectral axis}\label{sec:early-late-comparison}

We first test directional memory directly by comparing the early spectral axis with the late axis of the same trajectory. The observed early value is $Q_{2,\mathrm{obs}}=Q_2(30)$. The late value is the complex time average
$Q_{2,\mathrm{late}}=\langle Q_2(t)\rangle_{340\leq t\leq400}$ over the $31$ stored times in that window. If the axis is steady during the window, the complex average retains that axis. If it changes, contributions from different directions partly cancel, reducing the magnitude of the average relative to the mean of the magnitudes. We measure this cancellation with

\begin{equation}\label{eq:coherence}
\Gamma_{\mathrm{late}}=
\frac{|\langle Q_2\rangle_{340\leq t\leq400}|}
{\langle|Q_2|\rangle_{340\leq t\leq400}}.
\end{equation}

This ratio is one for a constant axis and decreases when the axis changes during the late window. Its distribution and the check for preferred grid directions are reported in Appendix~\ref{app:C}.

We compare the axis at time $t$ with the late axis through the cosine of twice their angular difference. Within condition $c$, the amplitude-weighted average is

\begin{equation}\label{eq:7}
C_{w,c}(t)=
\frac{\sum_{i\in c}\operatorname{Re}[Q_{2,i}(t)Q_{2,\mathrm{late},i}^*]}
{\sum_{i\in c}|Q_{2,i}(t)||Q_{2,\mathrm{late},i}|}.
\end{equation}

The index $i$ labels the $48$ trajectories in the condition. Each term equals one for coincident axes and minus one for perpendicular axes after division by the two magnitudes. The amplitude weighting in Eq.~\eqref{eq:7} gives less influence to axes that are weakly defined. We average $C_{w,c}$ equally over the 27 conditions, preventing conditions with larger anisotropy from dominating. Equal trajectory weights are used for the separate comparison with the linear calculation below.

\subsection{Directional alignment and coarsening stage}\label{sec:direction-scale-comparison}\label{sec:stage-comparison}

To determine whether a larger amount of coarsening accompanies a larger directional change, we compare both quantities between pairs of saved times. In Eq.~\eqref{eq:7}, replacing $t$ by $t_1$ and $Q_{2,\mathrm{late}}$ by $Q_2(t_2)$ gives $C_{w,c}(t_1,t_2)$. Lower values of this alignment indicate a larger change in direction. For the simultaneous scale change, we calculate $\Delta\ln L_{1,c}(t_1,t_2)=\operatorname{median}_{i\in c}\ln[L_{1,i}(t_2)/L_{1,i}(t_1)]$. Here $t_1$ is the earlier observation time and $t_2-t_1$ is the elapsed time.

The simultaneous change $\Delta\ln L_{1,c}(t_1,t_2)$ does not by itself place different conditions at the same coarsening stage. To separate the effect of quench depth from a difference in coarsening stage, we instead compare the conditions at similar changes in characteristic length from the observation time to the late averaging window. All 27 parameter combinations are linearly unstable, but Eq.~\eqref{eq:2} shows that changing $r$ or $\bar\phi$ changes the growth rate. To quantify this difference, we estimate the time required for the initial noise amplitude $0.02$ to grow by a factor of 50 at the maximum linear rate:

\begin{equation}\label{eq:amplification}
\sigma_{\max}=\frac{M(r+3\bar\phi^2)^2}{4\kappa},
\qquad t_{\mathrm{amp}}=\frac{\log(1/0.02)}{\sigma_{\max}}.
\end{equation}

This estimate is approximately $63$--$296$, $16$--$29$, and $7$--$10$ for the shallow ($r=-0.5$), intermediate ($r=-1$), and deep ($r=-1.5$) quenches. It describes initial linear amplification rather than a measured saturation time. These amplification times show why $t=30$ can correspond to different amounts of structural evolution. At later times, the growing domains also become limited by the box size. Comparing stages of coarsening therefore requires a measure of how far the structures have evolved. We use the measured change in characteristic length for that purpose.

For a time $t$ before the late window, we quantify the remaining scale change by

\begin{equation}\label{eq:8}
\begin{aligned}
s_c(t)&=\operatorname{median}_{i\in c}\log\frac{L_{1,\mathrm{late},i}}{L_{1,i}(t)},\\
L_{1,\mathrm{late},i}&=\langle L_{1,i}(t)\rangle_{340\leq t\leq400}.
\end{aligned}
\end{equation}

Unlike $\Delta\ln L_{1,c}(t_1,t_2)$, which measures the scale change during a selected two-time interval, $s_c(t)$ measures the change still remaining from $t$ to the late window. A positive $s_c(t)$ indicates that, across trajectories in condition $c$, the characteristic length at $t$ is typically below its late-window mean. To compare conditions at the same selected value of $s_c$, we use the latest time at which each $s_c(t)$ curve crosses that value. The latest crossing avoids using an earlier transient in which the initially broad spectrum is still reorganizing. Appendix~\ref{app:C} gives the uncertainty procedure, alternative weights, and sensitivity to the crossing rule.

\subsection{Linear calculation of the early axis}\label{sec:linear-spectral-evolution}

To determine whether isotropic linear growth can select the observed early axis, we propagate each trajectory's own initial spectrum without using its later state. A finite initial field contains unequal power in different directions even though the ensemble and the dynamics are isotropic. Moreover, different wave-number shells in the same field need not favor the same axis. Linear growth can therefore select an axis by changing the relative weights of those shells. In the continuum linearized dynamics, $Q_2$ evolves as

\begin{equation}\label{eq:5}
Q_{2,\mathrm{lin}}(t)=
\frac{\sum_{\mathcal K}[S(\mathbf k,0)/|\mathbf k|]e^{2\sigma(k)t}e^{2i\theta_{\mathbf k}}}
{\sum_{\mathcal K}[S(\mathbf k,0)/|\mathbf k|]e^{2\sigma(k)t}}.
\end{equation}

For a quantity $f_{\mathbf k}$ on the measured modes, write $\langle f\rangle_p=\sum_{\mathcal K}p_{\mathbf k}f_{\mathbf k}$. Differentiating the normalized weights gives

\begin{equation}\label{eq:6}
\frac{dQ_{2,\mathrm{lin}}}{dt}
=2\left[\langle\sigma e^{2i\theta}\rangle_p-\langle\sigma\rangle_p\langle e^{2i\theta}\rangle_p\right].
\end{equation}

\begin{figure}[H]
\centering
\includegraphics[width=\linewidth]{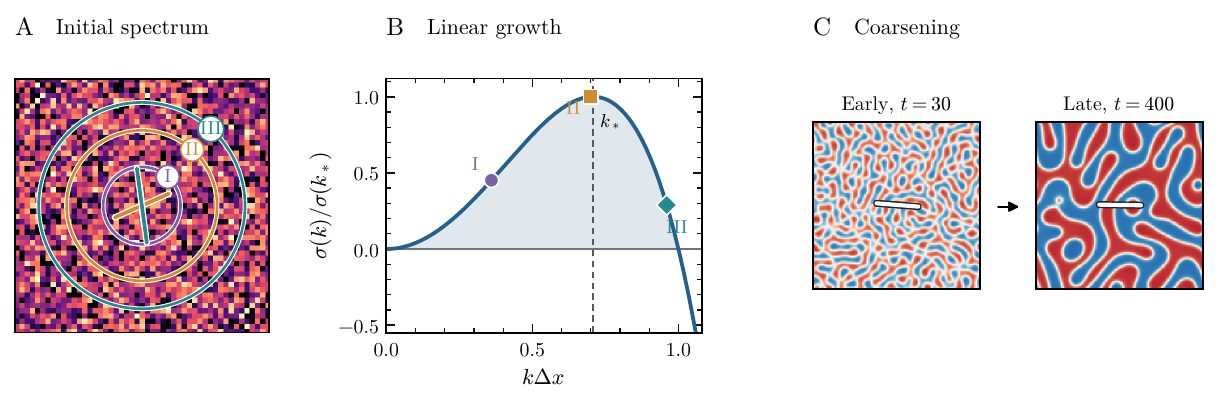}
\caption{\textbf{How linear growth selects the early spectral axis.} (A) Fourier intensity of one initial field. The three colored wave-number shells favor different axes, marked by the corresponding line segments. (B) Normalized linear growth rate $\sigma(k)/\sigma(k_*)$. The markers identify the shells from (A), and the dashed line marks the fastest-growing wave number $k_*$. Their different growth rates change the relative shell contributions to the full-spectrum axis. (C) The same field at $t=30$ and $t=400$, with white lines marking its spectral axes.}
\label{fig:linear-selection}
\end{figure}

Equation~\eqref{eq:6} says that the full-spectrum axis changes when shells with different growth rates also have different angular imbalances. If the normalized angular average is the same in every shell, radial amplification leaves $Q_2$ unchanged. Otherwise the magnitude and angle of the weighted sum over shells can change even though no individual wave vector rotates. This calculation explains how the initial spectrum can select an early axis; it is not used as a model of late nonlinear coarsening. Figure~\ref{fig:linear-selection} summarizes the shell reweighting and shows the early and late axes of the same field.

For the quantitative comparison at $t=30$, we linearize the same time-stepping rule that generates the trajectories. One numerical time step multiplies the Fourier amplitude by $g(k)$, so

\begin{equation}\label{eq:9}
g(k)=\frac{1-\Delta t\,M a k^2}{1+\Delta t\,M\kappa k^4},
\qquad
S_{\mathrm{lin}}(\mathbf k,t)=S(\mathbf k,0)|g(k)|^{2t/\Delta t}.
\end{equation}

We propagate the measured initial spectrum of each trajectory to $t=30$ and evaluate $Q_{2,\mathrm{lin}}$ with exactly the same Fourier range and weights as for $Q_{2,\mathrm{obs}}$. This supplies a trajectory-specific linear estimate of the observed early axis without fitting to the late pattern.

\subsection{Control for error in the linear calculation}\label{sec:linear-error-control}

To interpret the observed-minus-linear alignment difference, we first determine how much of it is caused by error in the calculated early axis. The observed early axis may align with the late axis more closely than the linear estimate simply because the calculation does not reproduce the observed $t=30$ axis exactly. Appendix~\ref{app:C} gives an idealized calculation of that effect. The main test below instead uses the measured angular errors directly.

For that test, we represent an axis by the unit complex number $u=Q_2/|Q_2|=e^{2i\psi}$, using the subscripts obs, lin, and late for the observed early, calculated early, and late axes. For trajectory $\ell$, we represent the angular error of the linear calculation by $\varepsilon_\ell=u_{\mathrm{lin},\ell}u_{\mathrm{obs},\ell}^*$. We apply this measured error to the observed early axis of a different trajectory $i$ from the same condition, giving $u_i^{(\ell)}=u_{\mathrm{obs},i}\varepsilon_\ell$. This exchange retains the dependence of the calculation error on $r$, $N$, and $\bar\phi$, while removing any relation between a trajectory's own calculation error and its own late axis. With $n_c=48$ and with $c(i)$ denoting the condition containing trajectory $i$, the average over the other 47 errors is

\begin{equation}\label{eq:11}
\begin{aligned}
\bar\varepsilon_{-i,c}&=\frac{\sum_{\ell\in c}\varepsilon_\ell-\varepsilon_i}{n_c-1},\\
\mathbb E_{\ell\ne i\mid c(i)}\!\left[\operatorname{Re}(u_i^{(\ell)}u_{\mathrm{late},i}^*)\right]
&=\operatorname{Re}[u_{\mathrm{obs},i}\bar\varepsilon_{-i,c(i)}u_{\mathrm{late},i}^*].
\end{aligned}
\end{equation}

Averaging the right-hand side of Eq.~\eqref{eq:11} over all trajectories gives the alignment $C_{\mathrm{exch},\mathrm{late}}$ produced by exchanging the measured errors. We compare the observed-minus-linear difference, $\Delta C_{\mathrm{raw}}=C_{\mathrm{obs},\mathrm{late}}-C_{\mathrm{lin},\mathrm{late}}$, with the difference caused by exchanged errors, $\Delta C_{\mathrm{err}}=C_{\mathrm{obs},\mathrm{late}}-C_{\mathrm{exch},\mathrm{late}}$. Their difference is

\begin{equation}\label{eq:residual_identity}
\Delta C_{\mathrm{res}}=\Delta C_{\mathrm{raw}}-\Delta C_{\mathrm{err}}
=C_{\mathrm{exch},\mathrm{late}}-C_{\mathrm{lin},\mathrm{late}}.
\end{equation}

If $\Delta C_{\mathrm{res}}$ is consistent with zero, the measured angular error of the linear estimate accounts for the observed-minus-linear difference. Appendix~\ref{app:C} gives the resampling, weighting, and checks that restrict error exchange to more closely matched trajectories.

\subsection{Prediction from early measurements}\label{sec:prediction-methods}

To test whether the early state contains trajectory-specific information about the late state, we ask whether it reduces prediction error beyond a baseline that knows only $r$, $N$, and $\bar\phi$. Prediction models connect early phase-field states with later microstructures \cite{Zapiain2021,Oommen2022,Rieger2024}. The baseline uses only the condition $c=(r,N,\bar\phi)$. For a scalar target, it predicts the median of the training values in that condition. For $Q_{2,\mathrm{late}}$, it predicts the geometric median of the complex training values in that condition: the point in the complex plane that minimizes the sum of distances to them \cite{VardiZhang2000}. For secondary comparisons of late morphology, the scalar targets are the late-window averages of $L_1$, $A_S$, and $\rho_{\chi,-}$. Here $\rho_{\chi,-}$ is the Euler characteristic of the $\phi\leq0$ phase divided by the system area; the Euler characteristic counts connected regions minus holes. Its discrete implementation and the other scalar measurements are defined in Appendix~\ref{app:A} \cite{Michielsen2001}.

Models use early measurements to predict trajectories excluded from fitting. We divide the data into five parts, fit on four, and test on the fifth, repeating this division five times. Any adjustable setting is chosen using only the four fitting parts. Consequently, no reported test trajectory helps to fit or select its own model. Appendix~\ref{app:D} gives the exact partitions, input sets, algorithms, and settings.

For target $j$, we define the mean absolute error (MAE) and the MAE-based skill score, which gives the fractional MAE reduction relative to the condition baseline:

\begin{equation}\label{eq:mae-skill-score}
\operatorname{MAE}_{m,j}=\frac1{1296}\sum_i|\widehat y^{(m)}_{ij}-y_{ij}|,
\qquad
\mathrm{SS}_{\mathrm{MAE},j}=1-\frac{\operatorname{MAE}_{\mathrm{model},j}}
{\operatorname{MAE}_{\mathrm{condition},j}}.
\end{equation}

Here $m$ labels the prediction method, and $y_{ij}$ and $\widehat y_{ij}^{(m)}$ are the observed and predicted target values. The modulus is an absolute value for a scalar and a Euclidean distance in the complex plane for $Q_{2,\mathrm{late}}$. Positive $\mathrm{SS}_{\mathrm{MAE}}$ therefore means that an early measurement reduces MAE beyond the known condition. All baselines, transformations, and model choices are estimated without the trajectories on which the error is evaluated \cite{VarmaSimon2006,KapoorNarayanan2023}.

Alongside the general prediction models, we use a direct test in which the early complex value enters only through a nonnegative real multiple:

\begin{equation}\label{eq:complex-persistence-map}
\widehat Q_{2,\mathrm{late}}=b_c+\lambda Q_2(30),\qquad \lambda\geq0.
\end{equation}

Here $b_c$ is the training-set condition baseline. The slope $\lambda$ is either shared by all conditions or fitted separately in each condition. A zero slope recovers the baseline; a positive real slope adds a correction along the observed early axis. We compare the full complex value $Q_2(30)$, which contains both magnitude and axis, with two controls: its magnitude alone and the same magnitude combined with the axis of a different trajectory in the same condition. These comparisons determine whether any reduction in prediction error requires the early direction to remain paired with its own late state. Details of the regression models, scalar controls, and uncertainty intervals are given in Appendix~\ref{app:D}.

\section{Results}\label{sec:results}

\subsection{Spectral selection and coarsening dynamics}\label{sec:spectral-coarsening-results}

\begin{figure}[H]
\centering
\includegraphics[width=\linewidth]{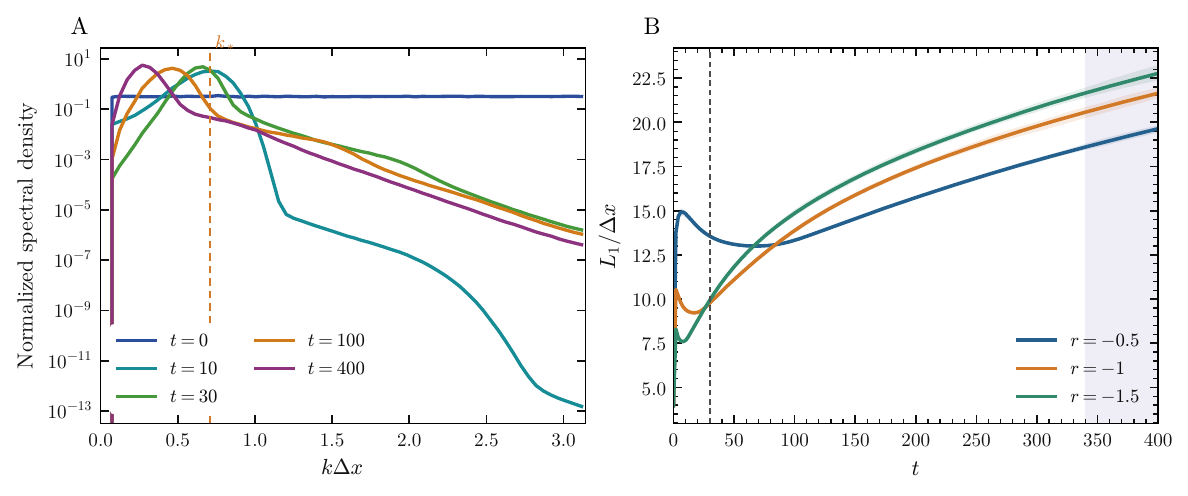}
\caption{\textbf{Radial selection and subsequent scale growth.} (A) Normalized radial spectra for $48$ trajectories at $N=128$, $\bar\phi=0$, and $r=-1$. The dashed line marks the fastest-growing wave number $k_*$ of the continuum linearized equation. (B) Mean characteristic length $L_1(t)$ for the three quenches, in units of $\Delta x$. Bands contain the middle $50\%$ of trajectories; the vertical line marks $t=30$, and shading marks the late averaging window.}
\label{fig:spectral-scales}
\end{figure}

We first verify the sequence of wavelength selection and domain growth described in Sec.~\ref{sec:model-setup}. Figure~\ref{fig:spectral-scales}(A) shows the normalized radial spectra for the $48$ trajectories at $N=128$, $\bar\phi=0$, and $r=-1$. The initially broad spectrum develops a band near the fastest-growing wave number predicted by the linear calculation. At later times, its peak moves toward smaller wave numbers as the domains grow. Normalization removes the increase in overall spectral power associated with growing composition contrast, making the redistribution across wave numbers visible.

Figure~\ref{fig:spectral-scales}(B) shows the corresponding characteristic length $L_1(t)$ for all three quench depths. During the initial redistribution of spectral power, $L_1$ can decrease even while the composition fluctuations grow. It then increases as coarsening becomes dominant. The three quenches pass through this sequence at different rates and select different early wavelengths. Consequently, comparing their spectral axes at a fixed time also compares patterns at different stages of coarsening. The following subsection first measures early--late alignment and then accounts for the difference in how far the quenches have coarsened.

\subsection{Directional alignment during coarsening}\label{sec:directional-age-results}

\begin{figure}[H]
\centering
\includegraphics[width=\linewidth]{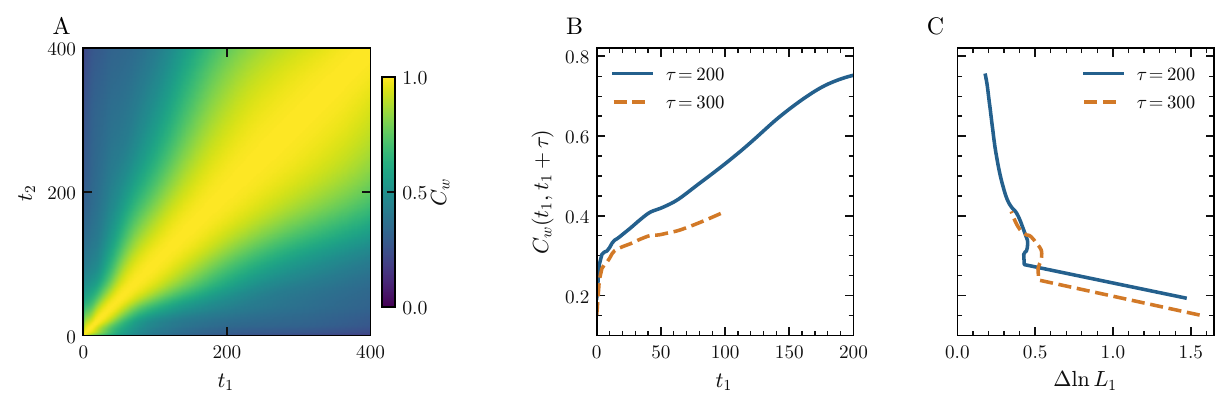}
\caption{\textbf{Directional alignment between two times during coarsening.} All panels show $C_w$, the alignment $C_{w,c}$ from Eq.~\eqref{eq:7} averaged equally over the $27$ conditions. (A) Alignment at all pairs of the $201$ stored times. (B) Alignment for fixed elapsed intervals $\tau=t_2-t_1=200$ and $300$, plotted against $t_1$. (C) The same alignment plotted against the logarithmic scale change $\Delta\ln L_1$, obtained by averaging the within-condition medians $\Delta\ln L_{1,c}$ equally. Lines follow increasing $t_1$.}
\label{fig:two-time}
\end{figure}

The radial spectra in Fig.~\ref{fig:spectral-scales} establish that the patterns change substantially between the early and late observations. We now ask whether the spectral axis measured during this evolution remains related to the later axis of the same trajectory. Figure~\ref{fig:two-time}(A) compares axes measured at two saved times, $t_1$ and $t_2$, using the alignment in Eq.~\eqref{eq:7}. The diagonal is unity because it compares an axis with itself. Away from the diagonal, the alignment remains larger when the two observation times are later and closer together.

Panels B and C make this trend quantitative for elapsed intervals $t_2-t_1=200$ and $300$. For an interval of $200$, the alignment averaged equally over conditions rises from about $0.19$ to $0.75$ as the earlier observation time $t_1$ increases. As $t_1$ is moved to later times, the logarithmic change in characteristic length, $\Delta\ln L_1$, falls from about $1.46$ to $0.18$. Here the scale change is first summarized by its median within each condition and then averaged equally over conditions. The interval of $300$ follows the same trend. Over the same elapsed interval, later observations therefore show both a smaller relative scale change and closer directional alignment. The amount of coarsening thus helps distinguish directional changes over equal elapsed intervals.

\begin{figure}[H]
\centering
\includegraphics[width=\linewidth]{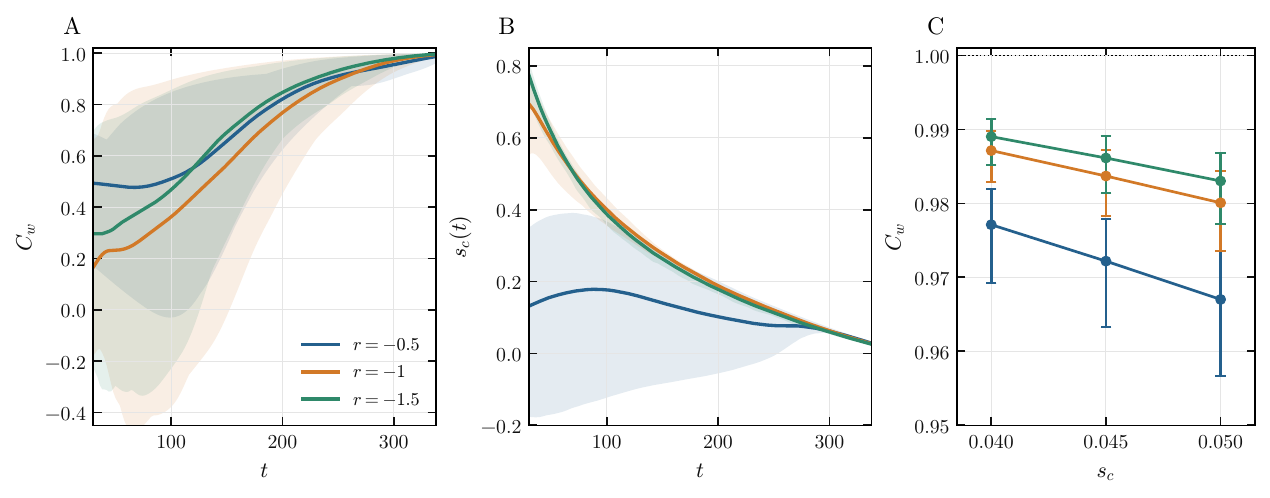}
\caption{\textbf{Directional alignment at a fixed time and after comparable changes in scale.} (A) Alignment $C_w$ between the axis at time $t$ and the mean axis over $340\leq t\leq400$. (B) Remaining logarithmic scale change $s_c(t)$ from Eq.~\eqref{eq:8}. In (A) and (B), lines are equally weighted means and bands span the nine condition curves in each quench. (C) Alignment at three $s_c$ values covered by every condition, with the vertical scale expanded near unity. The latest crossing of each selected value supplies the comparison time for each condition; bars are paired $95\%$ uncertainty intervals obtained by resampling trajectories.}
\label{fig:dynamical-age}
\end{figure}

We next compare the axis at time $t$ with the mean axis over $340\leq t\leq400$ in the same trajectory. Figure~\ref{fig:dynamical-age}(A) shows this alignment throughout the observation interval. At $t=30$, the amplitude-weighted alignments in Eq.~\eqref{eq:7}, averaged equally over conditions within each quench, are $0.493$, $0.165$, and $0.296$ for the shallow, intermediate, and deep quenches, respectively. These positive values describe a relation between the early and late axes within individual trajectories; the ensemble itself has no preferred laboratory direction. The early axis thus remains associated with the later pattern while its characteristic scale continues to change.

The unequal values at $t=30$ must be interpreted together with the unequal scale evolution established above. Figure~\ref{fig:dynamical-age}(B) plots the remaining logarithmic change from $L_1(t)$ to the late-window mean, as defined in Eq.~\eqref{eq:8}. At $t=30$, the shallow quench is much closer to its late characteristic length than the intermediate and deep quenches. The stronger early-to-late alignment of the shallow quench is therefore accompanied by a smaller change in scale between the early and late observations.

To separate this timing difference from quench dependence, Fig.~\ref{fig:dynamical-age}(C) compares the conditions at similar remaining changes in $L_1$. All $27$ condition curves cover the narrow interval $0.031\leq s_c\leq0.057$, which lies close to the late averaging window. At three selected values in this interval, the latest crossing of each condition curve occurs near $t\simeq315$--$320$. The alignment exceeds about $0.97$ for every quench under the weighting in Eq.~\eqref{eq:7}, and the shallow quench no longer shows the much stronger alignment seen at $t=30$. The differences at the fixed early time therefore partly reflect how far the three quenches have progressed in scale by that time. The complete $27$ condition curves, alternative weights, sensitivity to the crossing choice, and tabulated alignment values are reported in Appendix~\ref{app:C}.

These comparisons show that the early axis remains related to the late axis of the same trajectory, and that the strength of this relation changes with the accompanying growth of the characteristic length. They do not yet determine how the early axis was selected. We address that question by comparing the observed axis at $t=30$ with the axis obtained by linearly propagating the initial spectrum.

\subsection{Comparison with linear evolution}\label{sec:linear-reference-comparison}

\begin{figure}[H]
\centering
\includegraphics[width=0.95\linewidth]{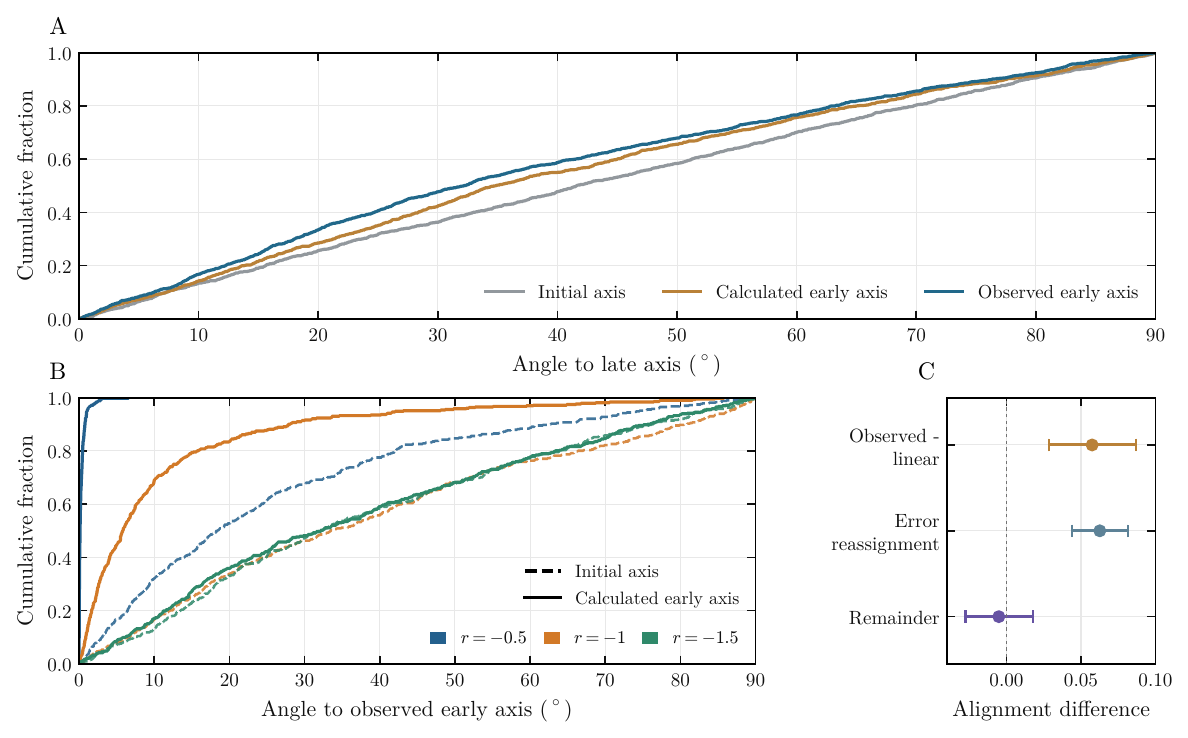}
\caption{\textbf{The calculated early axis and its difference from the observed axis.} (A) Cumulative distributions of the angles between the initial, calculated $t=30$, and observed $t=30$ axes and the mean late axis over $340\leq t\leq400$ in the same trajectory. (B) Cumulative distributions of the angular differences from the observed $t=30$ axis. Dashed lines use the initial axis and solid lines the calculated early axis; colors identify quench depth. (C) The observed-minus-linear alignment difference, the difference produced by reassigning the measured calculation errors within each condition, and the remainder after subtraction. Bars are paired $95\%$ uncertainty intervals obtained by resampling trajectories.}
\label{fig:linear-reference}
\end{figure}

The linear calculation in Eq.~\eqref{eq:9} uses each trajectory's measured initial spectrum and contains no information from its later pattern. Figure~\ref{fig:linear-reference}(A) compares the initial, calculated $t=30$, and observed $t=30$ axes with the late axis of the same trajectory. To understand this comparison with the late axis, we first check how well linear propagation reproduces the observed early axis, using the angular errors in Fig.~\ref{fig:linear-reference}(B). With the condition averaging and amplitude weighting of Eq.~\eqref{eq:7}, the alignment with the observed $t=30$ axis increases from approximately $0.46$ for the initial axis to approximately $0.76$ for the calculated $t=30$ axis. Panel B shows the remaining angular error of the calculation, which is larger for deeper quenches because $t=30$ then includes more growth and nonlinear rearrangement.

With equal trajectory weights, the observed $t=30$ axis is more closely aligned with its own late axis than the calculated $t=30$ axis is. The observed-minus-calculated difference is $0.058$, with interval $[0.028,0.087]$. This difference does not by itself show that nonlinear evolution turns the observed early axis toward the late axis. An angular error in the calculated early axis can itself reduce the alignment between that calculated axis and the late axis. An idealized calculation of that effect is given in Appendix~\ref{app:C}; the main control in Eq.~\eqref{eq:11} instead uses the measured angular differences directly.

Figure~\ref{fig:linear-reference}(C) shows the result of this control. The observed axis exceeds the calculated axis in late alignment by $0.0575$. Exchanging the measured calculation errors within each condition produces a difference of $0.0626$. Subtracting this control effect leaves a remainder of $-0.0051$, with interval $[-0.0275,0.0178]$. Thus the measured angular error in the calculated $t=30$ axis accounts for its alignment difference from the observed $t=30$ axis. Appendix~\ref{app:C} gives the contribution distributions, results under alternative weights and for individual quenches, and checks that restrict which errors may be reassigned.

The comparison assigns different roles to the successive parts of the evolution. Unequal angular power in a finite initial spectrum is reweighted during linear growth, and the linear calculation estimates the resulting early axis with finite accuracy. Nonlinear saturation and coarsening then transform the pattern while the observed $t=30$ axis remains related to its late axis. The angular error of the calculation explains why the observed early axis aligns more closely with the late axis than the calculated early axis does. A separate test is needed to determine whether measuring the early axis provides information about the later state of that particular trajectory.

\subsection{Prediction from early measurements}\label{sec:prediction-results}

Prediction supplies this separate test. The condition $c=(r,N,\bar\phi)$ already determines much of the late morphology, so every model using an early measurement is compared with the condition-only prediction defined in Sec.~\ref{sec:prediction-methods}. The models are fitted and evaluated on separate trajectories using the procedure described there; Appendix~\ref{app:D} gives the algorithms and validation details.

Appendix~\ref{app:D} quantifies how much of the late scalar variation the condition baseline already captures. The main comparison below asks whether the early spectral axis also adds information about the late complex $Q_2$.

\begin{figure}[H]
\centering
\includegraphics[width=\linewidth]{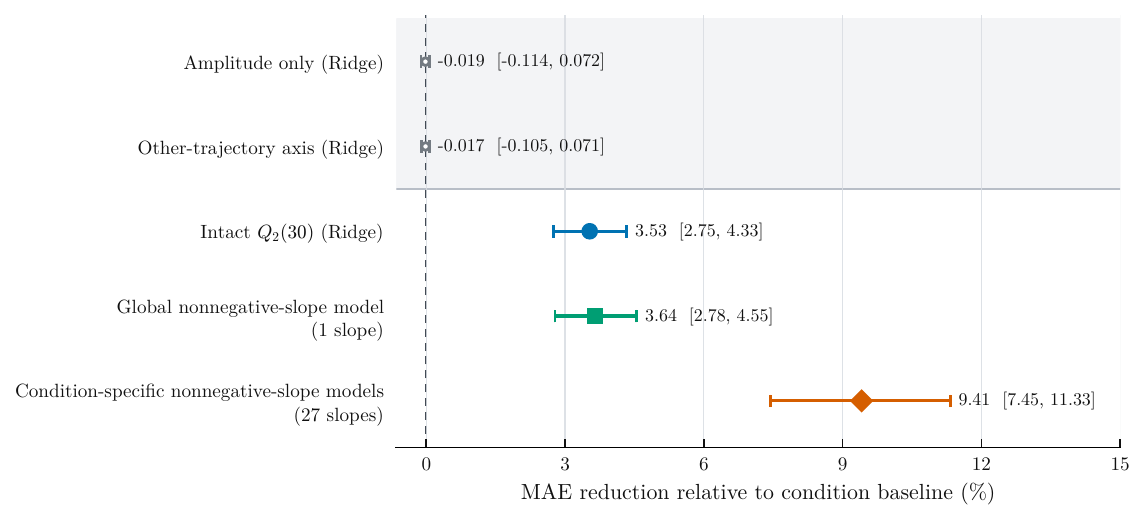}
\caption{\textbf{Reduction in late complex MAE supplied by early measurements.} Points show the percentage MAE reduction relative to the condition baseline, $100\mathrm{SS}_{\mathrm{MAE}}$ from Eq.~\eqref{eq:mae-skill-score}; bars are paired $95\%$ uncertainty intervals obtained by resampling trajectories. The first three rows compare Ridge inputs containing only $|Q_2(30)|$, that magnitude paired with another trajectory's axis from the same condition, and the full $Q_2(30)$. The last two rows use the nonnegative-slope model in Eq.~\eqref{eq:complex-persistence-map}, with one slope shared across conditions or one per condition. Values at or below zero do not improve on the baseline.}
\label{fig:prediction-mae-reduction}
\end{figure}

Figure~\ref{fig:prediction-mae-reduction} reports the central prediction test for the late complex quantity $Q_{2,\mathrm{late}}$, which contains both the strength and axis of the late spectral anisotropy. A regularized linear model using the real and imaginary parts of $Q_2(30)$ reduces complex MAE by $3.53\%$ relative to the condition baseline, with a $95\%$ interval of $[2.75,4.33]\%$ \cite{HoerlKennard1970}. A second model based on regression trees gives a comparable reduction \cite{Geurts2006}. The full $Q_2(30)$ therefore contains information about the late value that is absent from the condition alone.

Two controls locate this information in the early direction. Keeping only $|Q_2(30)|$ removes the measured axis, while combining that magnitude with the $t=30$ axis of another trajectory under the same condition breaks the pairing between the early and late axes of one trajectory. Neither control clearly reduces MAE because both intervals include zero. The improvement obtained from the full $Q_2(30)$ therefore requires the observed $t=30$ axis to remain paired with the late axis of the same trajectory.

The constrained linear map in Eq.~\eqref{eq:complex-persistence-map} tests the same relation by adding a correction along the observed early axis. One nonnegative slope shared across all conditions gives an MAE reduction similar to the regularized linear model. Allowing one nonnegative slope for each of the $27$ conditions reduces complex MAE by $9.41\%$ relative to the condition baseline, with interval $[7.45,11.33]\%$. Replacing each trajectory's $t=30$ axis with the axis of another trajectory in the same condition removes this gain. Adding the earlier complex observations at $t=0,6,14$ to $Q_2(30)$ provides no further benefit in the two tested model families. Appendix~\ref{app:D} reports the error distributions, sensitivity to the direction replacements, and paired model comparisons.

As a secondary check, early scalar histories provide no consistent improvement over the condition baseline for late $L_1$, $A_S$, and $\rho_{\chi,-}$. A separate model uses only the corresponding scalar value at $t=30$, with one nonnegative slope shared by all conditions. Among its three targets, only $A_S$ has an MAE-reduction interval entirely above zero: $0.683\%$, with interval $[0.053,1.293]\%$. Appendix~\ref{app:D} gives the input histories, methods, and complete scalar results.

The percentages for complex and scalar targets should not be compared as a single measure because their prediction errors have different meanings. For the measurements tested here, the observed spectral axis at $t=30$ improves prediction of the late complex anisotropy in the same trajectory, whereas the early scalar histories add little beyond what the condition already predicts. This independent prediction result supports the directional association found above without introducing a preferred direction shared across trajectories.

\section{Discussion}\label{sec:discussion}

Descriptions of spinodal decomposition usually emphasize what many realizations share: the growth of a characteristic length, the radially averaged structure factor, and scaling behavior. The present results show that a finite realization can also retain a directional record of its own. Measurements of $Q_2$ show that an axis selected from weak initial angular imbalances remains related to the late axis of the same trajectory after domain formation and coarsening. This persistence is compatible with isotropy because it concerns the relation between two times within one evolution. Different initial fields select different axes, so their individual directional histories do not imply a preferred laboratory direction across the ensemble.

Wave-number-dependent linear amplification provides the selection mechanism. At a fixed wave-number magnitude, the Cahn--Hilliard growth rate gives every direction the same amplification. Different wave-number shells in a finite initial field nevertheless have different angular imbalances. Their unequal growth changes their contributions to $Q_2$, allowing the spectral axis to move and strengthen even though no Fourier wave vector rotates. Equation~\eqref{eq:6} expresses this change in relative weights directly. The linear calculation recovers much of the early selection but does not reproduce the observed early axis exactly. Under the primary equal-trajectory weighting, the uncertainty interval for the remaining alignment difference includes zero after the measured angular errors are reassigned within conditions. The comparison therefore detects no additional nonlinear tendency that steers the early axis toward the late one. Nonlinear evolution transforms the morphology while retaining part of the selected directional information, even though the axis can continue to change.

Prediction tests whether the observed directional correlation carries information about the late state of an individual evolution. The condition baseline accounts for differences associated with quench depth, mean composition, and box size. The improvement obtained after adding the full early $Q_2$ therefore comes from the individual early observation. The controls identify which part of that observation matters: the magnitude alone gives no resolved improvement, and a direction borrowed from another trajectory under the same condition does not retain the gain. The early direction must remain paired with its own later state. Early scalar histories, by comparison, offer little improvement over their own condition baselines in the tested models. The early spectral axis thus carries information about its trajectory's later state beyond the tendencies set by the physical condition.

Comparisons across quench depths also show why the stage of coarsening matters when directional memory is measured. Quench depth changes both the selected wavelength and the rate of later growth, so patterns observed at a fixed time can be at different stages. A fixed-time comparison consequently combines differences between quenches with differences in how far their structures have evolved. The remaining logarithmic change in $L_1$ provides a way to compare them after accounting for this structural progress. Within the range shared by all conditions, comparing trajectories at similar remaining scale changes removes much of the stronger alignment seen for the shallow quench at the fixed early time. The stage of phase ordering should therefore accompany elapsed time when directional memory is compared across conditions, consistent with the established scaling description of conserved coarsening \cite{Bray1994,YeungRaoDesai1996,Furukawa1990}.

The scope of these conclusions is set by the observable, the geometry, and the dynamics. The measured $Q_2$ averages the Fourier intensity of the complete periodic pattern into one global axis and magnitude. It does not resolve local orientations or retain all the information in the intensity pattern. In the subset used for the numerical checks in Appendix~\ref{app:B:conservation}, the final characteristic length is approximately $0.14$--$0.38$ of the box side, illustrating the finite-system setting in which domains occupy an appreciable fraction of the available length. In a much larger system, one global $Q_2$ would average over many spatial regions and the early--late relation could become weaker. The simulations are two-dimensional, deterministic, and unforced after initialization, and they cover a finite coarsening window rather than an asymptotic regime. The present comparison does not establish how the relation changes in other geometries or under dynamics that can impose or rotate a direction.

The observable also has a direct experimental meaning. Because $Q_2$ is computed from Fourier intensity rather than spatial Fourier phase, it can be evaluated from time-resolved coherent-scattering patterns of phase-separating samples, of the kind used to follow speckle during domain coarsening \cite{Sutton1991,Malik1998,Brown1997,Brown1999}. Early and late intensity maps from repeated evolutions would therefore support the same comparison between correctly paired directions and directions reassigned across samples, with $L_1$ providing the corresponding measure of coarsening stage.

\section{Conclusion}\label{sec:conclusion}

In a finite sample drawn from an isotropic distribution, the initial composition fluctuations need not have equal power in every direction. Linear growth amplifies different wave numbers at different rates and selects an early spectral axis. Nonlinear domain formation and coarsening then transform the morphology, while the early and late axes of the same trajectory remain correlated. After accounting for the angular error of the linear calculation, the comparison detects no additional nonlinear steering tendency. Prediction and direction-reassignment controls further show that the correctly paired early axis carries information beyond that supplied by the physical condition. A fixed observation time corresponds to different amounts of coarsening across quenches, so matching the remaining scale change clarifies the comparison of directional persistence. In the finite, two-dimensional deterministic Cahn--Hilliard systems studied here, directional memory is therefore inherited from early spectral selection and persists through coarsening.

\section*{Data and code availability}
The simulation data and analysis code supporting this study are available on Zenodo at \href{https://doi.org/10.5281/zenodo.22633670}{10.5281/zenodo.22633670} \cite{DirectionalMemoryDataset}, under the Creative Commons Attribution 4.0 International (CC BY 4.0) license.

\appendix
\numberwithin{equation}{section}
\counterwithin{table}{section}
\counterwithin{figure}{section}
\renewcommand{\theequation}{\thesection\arabic{equation}}
\renewcommand{\thetable}{\thesection\arabic{table}}
\renewcommand{\thefigure}{\thesection\arabic{figure}}
\renewcommand{\theHequation}{appendix.\thesection.\arabic{equation}}
\renewcommand{\theHtable}{appendix.\thesection.\arabic{table}}
\renewcommand{\theHfigure}{appendix.\thesection.\arabic{figure}}

\section{Numerical methods and observable definitions}\label{app:A}

This appendix specifies the numerical update, the measurements made from each saved field, and the time intervals used to construct the prediction inputs and targets. These details make the main comparisons reproducible and distinguish properties of the evolving field from choices made only when it is measured.

\subsection{Evolution and stored fields}\label{app:A:evolution}

To specify exactly which dynamics generate the reported trajectories, we give the update rule and the treatment of the conserved mean. The periodic field is evolved with a semi-implicit Fourier scheme, following the established spectral approach to phase-field dynamics \cite{ChenShen1998,Zhu1999}. The gradient contribution to the chemical potential is implicit, while the local polynomial is evaluated from the current field. With an unnormalized forward discrete Fourier transform, the update is

\begin{equation}\label{eq:A1}
\widehat\phi^{n+1}_{\mathbf k}
=\frac{\widehat\phi^n_{\mathbf k}-\Delta t M k^2\operatorname{DFT}[r\phi^n+(\phi^n)^3]_{\mathbf k}}
{1+\Delta t M\kappa k^4}.
\end{equation}

Here $n$ is the time-step index and $\mathbf k$ is a Fourier-grid wave vector. All modes of the square Fourier grid enter this update. The circular support used for spectral measurements does not truncate the evolving field. The main trajectories evaluate the cubic term on the base grid without explicit dealiasing; Appendix~\ref{app:B} reports a paired comparison in which the cubic product is evaluated on a larger temporary grid. The main trajectories use $\Delta t=0.05$ and $\Delta x=1$, with double-precision evolution and single-precision saved fields. Appendix~\ref{app:B} states the altered settings used in the numerical comparisons.

\begin{figure}[H]
\centering
\includegraphics[width=\linewidth]{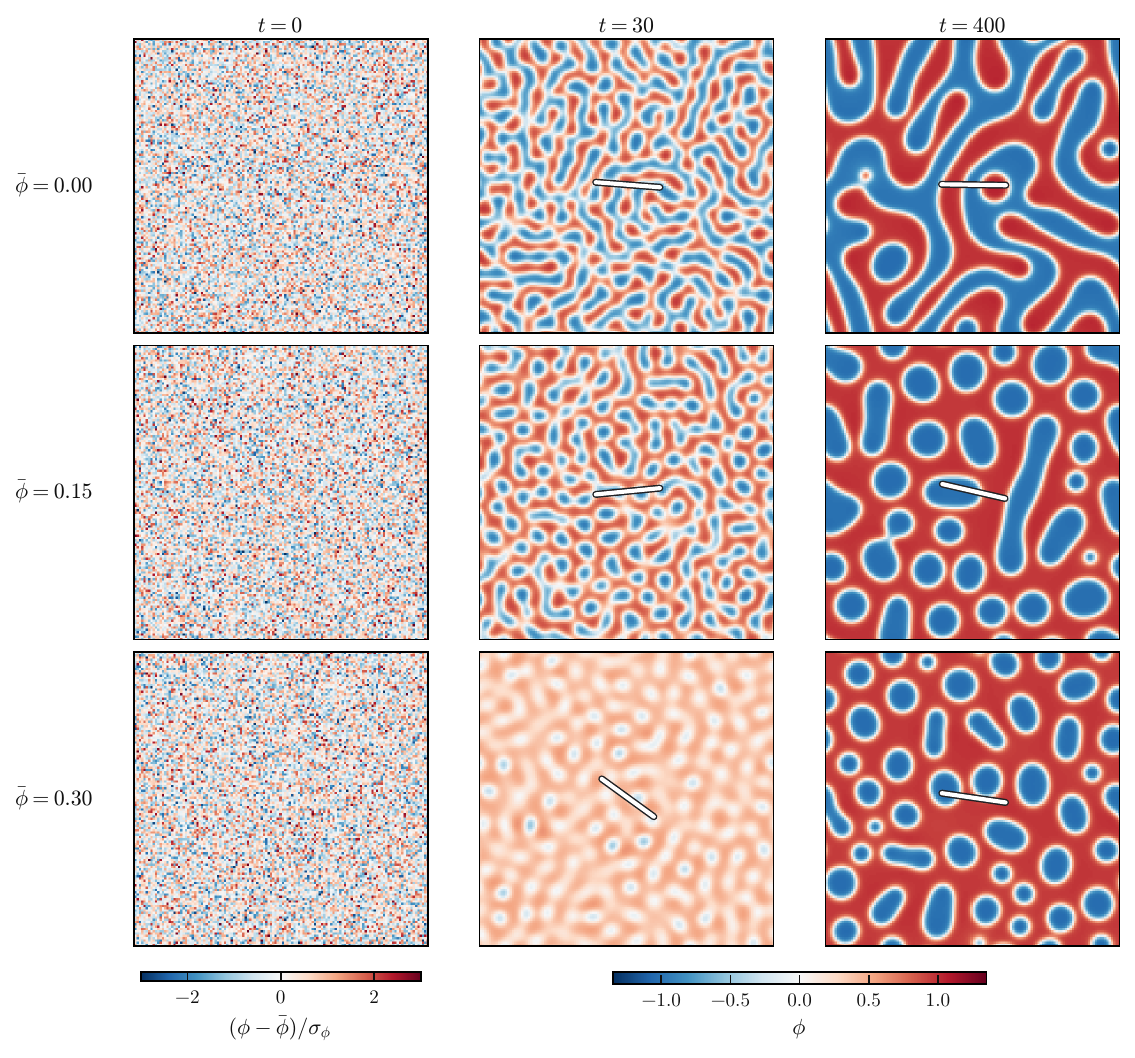}
\caption{Representative composition fields for three mean compositions. The $t=0$ panels show the prepared initial fluctuations, plotted as $(\phi-\bar\phi)/\sigma_\phi$. The $t=30$ and $400$ panels share the same composition scale $\phi$. White lines mark the spectral axes only at these two later times; their fixed length does not encode anisotropy magnitude.}
\label{fig:A:morphology}
\end{figure}

The zero Fourier mode is unchanged by Eq.~\eqref{eq:A1}, since its $k^2$ is zero. At every saving time, each field is projected to the prescribed spatial mean before conversion to single precision: $\phi\leftarrow\phi-\langle\phi\rangle_{\mathbf x}+\bar\phi$. The mass-conservation values in Appendix~\ref{app:B} therefore include evolution, projection, and storage. There are $201$ stored frames, separated by $2$ time units. The initial Gaussian noise has generator standard deviation $0.02$ and is then centered to the prescribed mean; its realized standard deviation is not reset to $0.02$ for every trajectory. Figure~\ref{fig:A:morphology} shows how representative fields at the three mean compositions develop from initial fluctuations into domains and then coarsen; the white lines identify the early and late spectral axes being compared.

\subsection{Spectral, topological, and local-field measurements}\label{app:A:observables}

To give each scalar and directional measurement an unambiguous numerical meaning, we state all conventions used to compute the ten features in Table~\ref{tab:A:features}. Measurements are evaluated on the saved fields converted to double precision. For a frame containing $N_{\mathrm{pix}}=N^2$ grid points, write its spatial mean as $\langle\phi\rangle_{\mathbf x}$ and its central moments as $m_\ell=N_{\mathrm{pix}}^{-1}\sum_{ij}(\phi_{ij}-\langle\phi\rangle_{\mathbf x})^\ell$. The spatial standard deviation is $\sigma_\phi=\sqrt{m_2}$, and the physical area is $\mathcal A=N^2\Delta x^2$. The weights $p_{\mathbf k}$ and circular support $\mathcal K$ are those in Sec.~\ref{sec:observables}; $\langle k^q\rangle_p=\sum_{\mathcal K}p_{\mathbf k}k^q$.

The topology measurement distinguishes connected regions from holes. It is computed from unions of foreground square cells on the periodic grid. For either phase, count its occupied faces $N_f$, distinct incident edges $N_e$, and distinct incident vertices $N_v$ with opposite boundaries identified. Then

\begin{equation}\label{eq:A2}
\chi_\pm=N_{v,\pm}-N_{e,\pm}+N_{f,\pm},
\qquad \rho_{\chi,\pm}=\frac{\chi_\pm}{\mathcal A}.
\end{equation}

Shared edges and vertices are counted once, including corner contacts and connections through periodic boundaries. The positive foreground is $\phi>0$ and the negative foreground is its complement, $\phi\leq0$. These rules specify how touching cells contribute to the discrete Euler characteristic \cite{Michielsen2001}.

For the interface density listed in Table~\ref{tab:A:features}, standardize each frame as $z=(\phi-\langle\phi\rangle_{\mathbf x})/\sigma_\phi$ and use periodic central differences,

\begin{equation}\label{eq:A:central}
D_x^{\mathrm{cen}} z_{ij}=\frac{z_{i+1,j}-z_{i-1,j}}{2\Delta x},
\qquad D_y^{\mathrm{cen}} z_{ij}=\frac{z_{i,j+1}-z_{i,j-1}}{2\Delta x}.
\end{equation}

The interface density integrates the gradient magnitude of a continuous standardized field; it is not the perimeter of a thresholded mask. The nearest-neighbor squared-difference measure instead uses the unscaled forward differences $\delta_x^+\phi_{ij}=\phi_{i+1,j}-\phi_{ij}$ and $\delta_y^+\phi_{ij}=\phi_{i,j+1}-\phi_{ij}$. It averages the squared forward differences in the two grid directions. The corresponding discrete gradient-energy density is obtained by multiplying by $\kappa/(2\Delta x^2)$; the main trajectories have $\Delta x=1$.

\begingroup
\small
\renewcommand{\arraystretch}{1.16}
\begin{longtable}{@{}>{\raggedright\arraybackslash}p{0.25\textwidth}>{\raggedright\arraybackslash}p{0.70\textwidth}@{}}
\caption{Operational definitions of the ten scalar features. All spatial indices wrap periodically. Spectral moments use the circular support and $p_{\mathbf k}\propto S(\mathbf k)/k$.}\label{tab:A:features}\\
\toprule
\textbf{Feature} & \textbf{Definition and convention}\\\midrule
\endfirsthead
\multicolumn{2}{l}{\small Table~\thetable\ continued}\\
\toprule
\textbf{Feature} & \textbf{Definition and convention}\\\midrule
\endhead
\bottomrule
\endfoot
Characteristic length & $L_1=2\pi/\langle k\rangle_p$, as in Eq.~\eqref{eq:4}.\\
Relative spectral width & $\sqrt{\langle k^2\rangle_p-\langle k\rangle_p^2}/\langle k\rangle_p$; a dimensionless width, not the variance alone.\\
Spectral anisotropy & $A_S=|Q_2|$, with $Q_2$ defined as in Eq.~\eqref{eq:4}.\\
Positive-phase fraction & $N_{\mathrm{pix}}^{-1}\sum_{ij}\mathbf 1_{\{\phi_{ij}>0\}}$; the threshold is zero.\\
Positive Euler density & $\rho_{\chi,+}=\chi_+/\mathcal A$ for the positive foreground in Eq.~\eqref{eq:A2}.\\
Negative Euler density & $\rho_{\chi,-}=\chi_-/\mathcal A$ for the complementary negative foreground in Eq.~\eqref{eq:A2}.\\
Interface density & $\mathcal A^{-1}\sum_{ij}\sqrt{(D_x^{\mathrm{cen}} z_{ij})^2+(D_y^{\mathrm{cen}} z_{ij})^2}\,\Delta x^2$; set to zero if $\sigma_\phi<10^{-12}$.\\
Field standard deviation & $\sigma_\phi=\sqrt{m_2}$, using the population spatial moment rather than an $N_{\mathrm{pix}}-1$ denominator.\\
Field skewness & $\dfrac{\sqrt{N_{\mathrm{pix}}(N_{\mathrm{pix}}-1)}}{N_{\mathrm{pix}}-2}\dfrac{m_3}{m_2^{3/2}}$, the bias-corrected Fisher--Pearson coefficient.\\
Nearest-neighbor squared-difference measure & $N_{\mathrm{pix}}^{-1}\sum_{ij}[(\delta_x^+\phi_{ij})^2+(\delta_y^+\phi_{ij})^2]$; the corresponding discrete gradient-energy density equals $\kappa/(2\Delta x^2)$ times this quantity.\\
\end{longtable}
\endgroup

\subsection{Temporal aggregation}\label{app:A:aggregation}

To ensure that every prediction comparison uses a precisely defined early record and late outcome, we specify how saved measurements are combined in time. The scalar input history concatenates the ten features in Table~\ref{tab:A:features} at $t=0,10,20,30$, followed by the ten differences between $t=30$ and $t=0$, yielding $50$ inputs. The directional histories instead use the real and imaginary parts of $Q_2$ at $t=0,6,14,30$, so the scalar and directional predictors use different sampling schedules. The late targets are arithmetic means over all $31$ stored frames at $340\leq t\leq400$. In particular, $Q_{2,\mathrm{late}}=\langle Q_2\rangle_{\mathrm{late}}$ and the scalar anisotropy target $\langle|Q_2|\rangle_{\mathrm{late}}$ are computed separately. Appendix~\ref{app:C} repeats the calculation with different late windows and radial weights using the same saved trajectories.

\section{Numerical robustness}\label{app:B}

This appendix tests whether the reported directional association could be produced by numerical error, limited resolution, spectral truncation, or the square simulation grid. All comparisons retain the early observation at $t=30$ and the late complex mean over $t=340$--$400$, and they distinguish changes in individual morphology from changes in ensemble alignment. The alignment comparisons require a corresponding linear calculation for each numerical treatment. When the grid or time step changes, the linear calculation uses that treatment's discrete update. The bandwidth and cubic-product comparisons share the same reference computed from the unfiltered initial field. Appendix~\ref{app:A} defines the baseline update and observables.

\subsection{Conservation, energy decay, and time step}\label{app:B:conservation}

We first check that the discrete evolution preserves the conserved mean, follows the expected free-energy decrease, and is insensitive to halving the time step. A balanced subset of $135$ trajectories from the main ensemble (five per condition) is used to evaluate the spatial mean and free energy of the saved fields. The largest absolute change in the spatial mean was $1.71\times10^{-9}$. None of the saved-step energy increases exceeded
\[
10^{-7}\max\!\left(1,\max_t\left|F(t)/L_{\mathrm{box}}^2\right|\right).
\]
These checks use the stored single-precision fields at intervals of $2$ time units, while evolution is performed in double precision. In this subset, the final ratios $L_1/L_{\mathrm{box}}$ range from $0.143$ to $0.378$: the characteristic domain length occupies an appreciable fraction of the box side.

The time-step comparison used nine matched pairs at $N=64$: three initial realizations for each of $(r,\bar\phi)=(-0.5,0),(-1,0.15),(-1.5,0.30)$. Each initial field was evolved to $t=400$ with $\Delta t=0.05$ and $0.025$, using the same saved times. The median symmetric relative difference in late $L_1$ was $4.93\times10^{-4}$, where a symmetric relative difference is $2|x-y|/(|x|+|y|)$. The difference between the late spectral-axis angles had median $1.26^\circ$ and $97.5$th percentile $4.75^\circ$. Table~\ref{tab:B:raw} reports the corresponding changes in the difference between the observed-to-late and linear-to-late alignments. This comparison spans three conditions; its paired intervals describe that sample rather than a time-step extrapolation over all $27$ conditions.

\subsection{Spatial refinement at fixed physical box size}\label{app:B:refinement}

To test whether spatial discretization changes the measured association, we refine the grid while holding the physical box size and the initial realization fixed. The fixed-box refinement contains four independently seeded matched pairs in each of the $27$ conditions ($108$ pairs total). The coarse grids used for the main ensemble, $(N,\Delta x)=(64,1),(96,1),(128,1)$, are paired with $(128,0.5),(192,0.5),(256,0.5)$, respectively. Thus each pair has the same physical box length. A coarse initial field is periodically Fourier-interpolated onto the fine grid, after which its mean and standard deviation are matched to the coarse field. The fine-grid member resolves the same band-limited realization; no additional independent high-frequency fluctuations are introduced. Both members use $\Delta t=0.05$ and the same observation times.

With equal trajectory weights, observed early-to-late alignment changes by $-0.00230$ (coarse minus fine), with $95\%$ paired interval $[-0.00719,0.00274]$. Table~\ref{tab:B:raw} reports the corresponding changes in the observed-minus-linear alignment difference under three common weighting choices. Individual fields agree to different degrees: the median and $97.5$th percentile of the late spectral-axis difference are $0.0885^\circ$ and $5.43^\circ$. For late $L_1$, the same percentiles of the symmetric relative difference are $0.0205\%$ and $0.904\%$; for late $|Q_2|$, they are $0.303\%$ and $29.5\%$. These small changes in ensemble alignment coexist with a broader relative-error tail in the late anisotropy amplitude.

\subsection{Spectral bandwidth and cubic-product evaluation}\label{app:B:bandwidth}

To determine whether unresolved high-wave-number interactions affect the result, we separately change the evolving spectral bandwidth and the grid used to evaluate the cubic product. The bandwidth comparison uses $54$ initial fields, two from each of the $27$ conditions. Each field is evolved under three treatments: unfiltered evolution and sharp componentwise cutoffs at $2/3$ and $1/2$ of the Nyquist wave number. Thus each treatment contains $54$ trajectories, paired through their common initial fields. In the two filtered treatments, both the nonlinear spectrum and the updated field spectrum are masked at every step. These treatments therefore remove field modes and also change their nonlinear interactions. All three treatments use the same unfiltered linear reference generated from their common initial field. Their observed-minus-linear alignment differences consequently compare the altered nonlinear histories with a fixed reference.

The separate cubic-product comparison also uses $54$ matched initial fields, with two fields per condition, and retains the full state bandwidth and the same unfiltered linear reference. At each step, normalized Fourier interpolation maps the current field onto a grid strictly larger than $2N$ in each direction. The cubic product is evaluated there and projected back to the base-grid spectrum; the $r\phi$ term remains on the base grid. The padded sizes are $132$, $196$, and $264$ for base sizes $64$, $96$, and $128$. Even-grid Nyquist coefficients are split and recombined during interpolation and projection. A separate implementation check pads $8\times8$ test fields to $18\times18$ and $25\times25$, both strictly larger than $2N$ in each direction. The retained cubic spectra agree to a relative precision of $1.11\times10^{-15}$. Across the three evolution grid sizes, the maximum padded-field imaginary-to-real root-mean-square ratio is $2.81\times10^{-15}$.

For these comparisons, $\Delta C_{\mathrm{raw}}=C_{\mathrm{obs},\mathrm{late}}-C_{\mathrm{lin},\mathrm{late}}$ denotes the observed-minus-linear alignment difference before error reassignment. The observed and linear terms use the same weight for each trajectory, so their difference compares the axes without also changing the weights. We use equal trajectory weights, late-amplitude weights $|Q_{2,\mathrm{late}}|$, and symmetric-amplitude weights $|Q_{2,\mathrm{late}}|\sqrt{|Q_{2,\mathrm{obs}}||Q_{2,\mathrm{lin}}|}$. Weights are normalized inside each physical condition, and the condition estimates are then averaged equally. Each numerical treatment uses its own measured amplitudes, while its observed and linear terms share those weights. All intervals in Table~\ref{tab:B:raw} use $10000$ paired bootstrap replicates within conditions, retaining the pairing between numerical treatments \cite{Efron1979}.

\begin{table}[htbp]
\centering
\caption{Paired changes in the condition-balanced observed-minus-linear alignment difference. Each entry is a point estimate followed by its $95\%$ paired-bootstrap interval. The comparison direction is stated in the first column. Spatial refinement has $108$ pairs; the time-step sample has nine pairs; each bandwidth or padding comparison uses $54$ initial fields evolved under both treatments being compared. These are changes in $\Delta C_{\mathrm{raw}}$, before the measured angular errors are reassigned.}
\label{tab:B:raw}
\small
\renewcommand{\arraystretch}{1.25}
\setlength{\tabcolsep}{4pt}
\begin{tabularx}{\textwidth}{@{}>{\raggedright\arraybackslash}X*{3}{>{\centering\arraybackslash}X}@{}}
\toprule
\textbf{Paired comparison} & \textbf{Equal trajectory} & \textbf{Late amplitude} & \textbf{Symmetric amplitude} \\
\midrule
$\Delta t=0.025$ minus $0.05$ & \shortstack{$0.0167$\\$[-0.0429,0.0497]$} & \shortstack{$0.0345$\\$[-0.0432,0.0503]$} & \shortstack{$0.0444$\\$[-0.0438,0.0503]$} \\
Coarse minus fine, fixed box & \shortstack{$0.0002$\\$[-0.0044,0.0049]$} & \shortstack{$-0.0032$\\$[-0.0256,0.0067]$} & \shortstack{$-0.0118$\\$[-0.0279,0.0069]$} \\
$2/3$ cutoff minus unfiltered & \shortstack{$0.0090$\\$[-0.0473,0.0672]$} & \shortstack{$0.0540$\\$[-0.0378,0.0844]$} & \shortstack{$0.0616$\\$[-0.0375,0.0850]$} \\
$1/2$ cutoff minus unfiltered & \shortstack{$-0.1295$\\$[-0.2364,-0.0230]$} & \shortstack{$-0.1042$\\$[-0.2247,-0.0073]$} & \shortstack{$-0.0983$\\$[-0.2234,-0.0035]$} \\
Strict padding minus base-grid cubic product & \shortstack{$0.0026$\\$[-0.0040,0.0090]$} & \shortstack{$0.0122$\\$[-0.0033,0.0178]$} & \shortstack{$0.0118$\\$[-0.0036,0.0183]$} \\
\bottomrule
\end{tabularx}
\end{table}

The $1/2$ cutoff reduces $\Delta C_{\mathrm{raw}}$ under all three weights. Under equal trajectory weights, its value changes from $0.1074$ in the matched unfiltered treatment to $-0.0222$. The $2/3$ cutoff gives $0.1164$, while strict cubic-product padding gives $0.1099$. Thus the stronger bandwidth cutoff changes the uncorrected observed-minus-linear difference much more than evaluating the cubic product on a larger temporary grid. Because the cutoff changes both the evolving modes and their nonlinear interactions, this comparison does not isolate the contribution of a particular wave-number band to directional memory. The paired intervals in Table~\ref{tab:B:raw} include zero for the $2/3$ cutoff and strict padding. The time-step, cutoff, and padding analyses reported here concern the alignment difference before error reassignment; the fixed-box comparison below directly tests the remaining difference after reassignment.

\subsection{Spatial refinement of the error-reassignment result}\label{app:B:residual}

To test whether spatial refinement changes the result after measured reference errors are reassigned, we apply the same within-condition procedure used in the main text to the coarse and fine members of the $108$ matched pairs. With common trajectory weights, Eq.~\eqref{eq:residual_identity} expresses the remaining difference $\Delta C_{\mathrm{res}}$ as the alignment obtained after reassignment minus the actual linear-to-late alignment. Every condition contains four pairs, so an error for each evaluated trajectory can be drawn from the three other pairs. The $10000$ bootstrap replicates resample four pair indices inside each condition, keep coarse and fine members together, and recompute both averages over reassigned errors in each draw.

\begin{table}[htbp]
\centering
\caption{Remaining alignment difference $\Delta C_{\mathrm{res}}$ after error reassignment in the $108$ fixed-box refinement pairs. Each entry contains a point estimate and its $95\%$ interval. Coarse and fine levels are condition-balanced; the last column is their paired difference. All three terms use the same trajectory weights within each numerical treatment.}
\label{tab:B:residual}
\small
\renewcommand{\arraystretch}{1.25}
\setlength{\tabcolsep}{4pt}
\begin{tabularx}{\textwidth}{@{}>{\raggedright\arraybackslash}X*{3}{>{\centering\arraybackslash}X}@{}}
\toprule
\textbf{Trajectory weight} & \textbf{Coarse} & \textbf{Fine} & \textbf{Coarse minus fine} \\
\midrule
Equal trajectory & \shortstack{$-0.0535$\\$[-0.1239,0.0429]$} & \shortstack{$-0.0569$\\$[-0.1270,0.0410]$} & \shortstack{$0.0034$\\$[-0.0026,0.0080]$} \\
Late amplitude & \shortstack{$-0.0260$\\$[-0.1181,0.0539]$} & \shortstack{$-0.0234$\\$[-0.1174,0.0599]$} & \shortstack{$-0.0026$\\$[-0.0260,0.0118]$} \\
Symmetric amplitude & \shortstack{$-0.0506$\\$[-0.1459,0.0485]$} & \shortstack{$-0.0386$\\$[-0.1404,0.0541]$} & \shortstack{$-0.0120$\\$[-0.0288,0.0132]$} \\
\bottomrule
\end{tabularx}
\end{table}

The paired changes in Table~\ref{tab:B:residual} are small relative to the uncertainty of either level, and all three intervals span zero. In particular, the remaining difference under equal-trajectory weighting changes by $0.00343$ with interval $[-0.00261,0.00801]$. This directly connects the refinement comparison to the main error-reassignment result. The broad level intervals reflect the four-pair condition samples and the three available error sources for each evaluated trajectory; the paired shift is the informative quantity for this comparison.

\subsection{Exact grid covariance and interpolated-rotation sensitivity}\label{app:B:rotation}

To separate exact symmetry of the square grid from errors introduced by interpolating a rotated field, we compare an exact grid rotation with an interpolated diagonal rotation. Rotation covariance here means that rotating an initial field and then evolving it gives the same result as evolving it first and then rotating the result. The dynamical rotation comparison starts from two independent initial realizations in each of the $27$ conditions, giving $54$ trajectories per treatment. The unrotated field, its exact $90^\circ$ rotation, and its $45^\circ$ rotation are separately evolved to $t=400$. The $45^\circ$ transformation uses cubic interpolation with periodic wrapping, followed by matching the original mean and standard deviation. Directional errors compare the resulting $Q_2$ value with the value expected from rotating the unrotated solution, using the spectral-axis distance modulo $180^\circ$.

\begin{table}[htbp]
\centering
\caption{Exact grid covariance and interpolated-rotation sensitivity for $54$ initial realizations, each evolved under the unrotated and two rotated treatments. Entries give the median and $97.5$th percentile of the spectral-axis error in degrees, rather than confidence intervals. Early values use $t=30$ and late values use the complex mean over $t=340$--$400$.}
\label{tab:B:rotation}
\small
\renewcommand{\arraystretch}{1.25}
\setlength{\tabcolsep}{4pt}
\begin{tabularx}{\textwidth}{@{}>{\raggedright\arraybackslash}X>{\raggedright\arraybackslash}X*{2}{>{\centering\arraybackslash}X}@{}}
\toprule
\textbf{Angle} & \textbf{Initial-field transform} & \textbf{Early error: median; upper percentile} & \textbf{Late error: median; upper percentile} \\
\midrule
$90^\circ$ & Exact grid rotation & \shortstack{$7.49\times10^{-14}$;\\$3.52\times10^{-12}$} & \shortstack{$8.28\times10^{-11}$;\\$1.35\times10^{-9}$} \\
$45^\circ$ & Periodic cubic interpolation & $9.89$; $71.39$ & $23.29$; $84.62$ \\
\bottomrule
\end{tabularx}
\end{table}

Exact $90^\circ$ rotation reproduces the covariance of the square grid, while the interpolated $45^\circ$ initial field produces appreciable trajectory-level differences. Table~\ref{tab:B:rotation} tests the response of individual trajectories to a rotated initial field. Figure~\ref{fig:C:direction-checks} instead describes laboratory-frame axes across realizations; these complementary comparisons distinguish trajectory covariance from ensemble directional isotropy.

\section{Robustness checks for directional alignment}\label{app:C}

This appendix tests whether the early-to-late directional association survives reasonable changes in measurement and comparison procedure. It varies the spectral definition and late window, checks for preferred grid directions, examines the reassignment of errors in the linear calculation, and compares quenches at similar stages of coarsening.

\subsection{Spectral weighting and the late observation window}\label{app:C:definition}

To rule out an association created by one radial weight or one choice of late window, we repeat the measurement under the alternatives below. The spectral axis describes an angular imbalance distributed over wave numbers, so its dependence on radial weighting can be tested without changing the underlying trajectories. With the same circular Fourier support as in the main analysis, define
\begin{equation}\label{eq:C:radial}
Q_2^{(\nu)}(t)=
\frac{\sum_{\mathbf k}k^{-\nu}S(\mathbf k,t)e^{2i\theta_{\mathbf k}}}
{\sum_{\mathbf k}k^{-\nu}S(\mathbf k,t)},
\qquad \nu=0,1,2.
\end{equation}
The zero mode is omitted. The primary definition is $\nu=1$; $\nu=0$ retains raw spectral power and $\nu=2$ gives more weight to small wave numbers. For each choice, both the early $Q_2$ value at $t=30$ and the late complex mean are recalculated. The late mean is taken before its direction is extracted. Table~\ref{tab:C:definition} crosses all three definitions with six late windows, including shifted windows of the same duration. Each entry is calculated with the amplitude-product weights of Eq.~\eqref{eq:7} within each of the $27$ conditions and then averaged equally over conditions. The numerical values change with the spectral definition, while the positive directional association remains throughout this set of choices.

\begin{table}[H]
\centering\small
\caption{Alignment of $Q_2(30)$ with its late complex mean under six late windows and three radial weights. These are point estimates for all $1296$ trajectories, not confidence intervals. All window endpoints are inclusive; stored frames are separated by $2$ time units. The primary window and weight are $340$--$400$ and $k^{-1}$.}\label{tab:C:definition}
\setlength{\tabcolsep}{5pt}
\begin{tabularx}{\linewidth}{@{}l r *{3}{>{\centering\arraybackslash}X}@{}}
\toprule
Late window & Frames & $k^0$ & $k^{-1}$ & $k^{-2}$\\
\midrule
$300$--$400$ & 51 & $0.3434$ & $0.3301$ & $0.2753$\\
$320$--$400$ & 41 & $0.3379$ & $0.3242$ & $0.2690$\\
$340$--$400$ & 31 & $0.3326$ & $0.3182$ & $0.2629$\\
$360$--$400$ & 21 & $0.3274$ & $0.3121$ & $0.2575$\\
$300$--$360$ & 31 & $0.3504$ & $0.3371$ & $0.2821$\\
$320$--$380$ & 31 & $0.3415$ & $0.3278$ & $0.2726$\\
\bottomrule
\end{tabularx}
\end{table}

\begin{figure}[H]
\centering
\includegraphics[width=\linewidth]{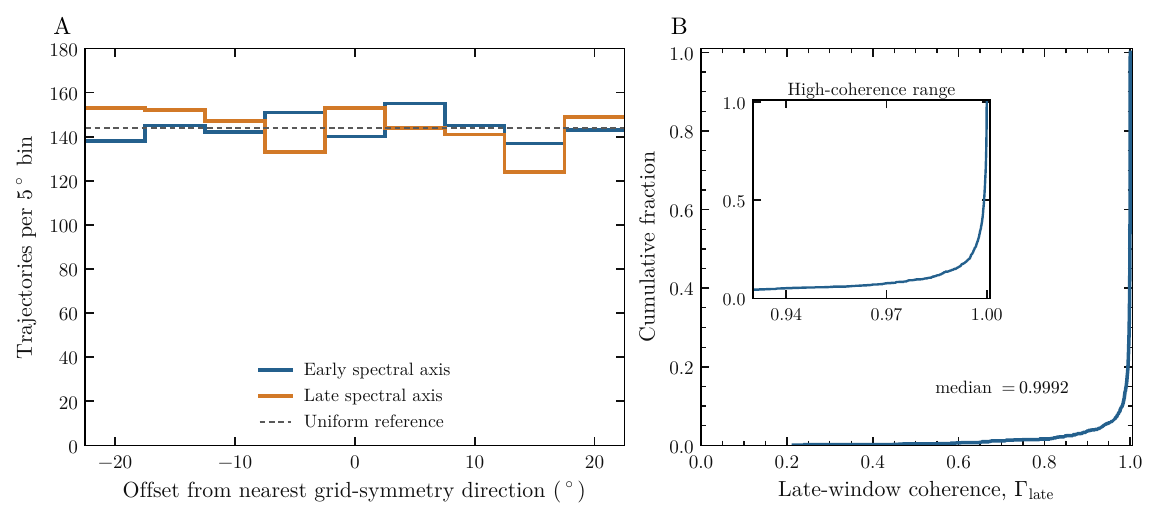}
\caption{Directional checks across the full ensemble. (A) Early and late spectral-axis distributions folded about the nearest axis or diagonal of the square grid, which together form its fourfold ($D_4$) symmetry directions. The dashed horizontal line gives the count for a uniform distribution. (B) Empirical cumulative distribution of late-window directional coherence for all $1296$ trajectories. The inset expands the region near unity, and the annotated value gives the median.}
\label{fig:C:direction-checks}
\end{figure}

We also check whether the late-window mean represents a stable direction within individual trajectories. The coherence $\Gamma_{\mathrm{late}}$ in Eq.~\eqref{eq:coherence} measures how consistently the instantaneous late axes point in the same direction; a value near one indicates little variation within the window. Panel B of Fig.~\ref{fig:C:direction-checks} shows its complete distribution: the median is $0.9992$ and the fifth percentile is $0.9372$. The square grid has fourfold ($D_4$) symmetry, so its axes and diagonals could in principle be preferred directions. Panel A verifies that the measured spectral axes do not concentrate along them. The magnitude of the late-axis ensemble mean, $|1296^{-1}\sum_i e^{2i\psi_{\mathrm{late},i}}|=0.0123$, is likewise small, confirming that the late axes have almost no preferred direction across trajectories.

\subsection{Common trajectory weights and error reassignment}\label{app:C:null}

We first show how angular error alone can lower the linear-to-late alignment. Define $X=2(\psi_{\mathrm{obs}}-\psi_{\mathrm{late}})$ and $\eta=2(\psi_{\mathrm{lin}}-\psi_{\mathrm{obs}})$. If $\eta$ is independent of $X$ and has a reflection-symmetric distribution, equal trajectory weighting gives

\begin{equation}\label{eq:10}
C_{\mathrm{obs},\mathrm{late}}-C_{\mathrm{lin},\mathrm{late}}
=C_{\mathrm{obs},\mathrm{late}}\bigl(1-C_{\mathrm{obs},\mathrm{lin}}\bigr),
\end{equation}

where $C_{A,B}$ is the mean of $\cos[2(\psi_A-\psi_B)]$, first over trajectories in each condition and then equally over conditions. This calculation is only a simple reference because its assumptions need not hold in the measured ensemble. We therefore make the main comparison by retaining every measured error and assigning it to other trajectories in the same physical condition.

The same fixed weight $w_i$ is used for trajectory $i$ in all three alignments: observed to late, linear to late, and reassigned-error to late. For the observed or calculated early axis, let $u_{Y,i}$ be its unit complex representation. For the reassigned-error comparison, let $u_{\mathrm{exch},i}=u_{\mathrm{obs},i}\bar\varepsilon_{-i,c(i)}$ be the complex average over all available errors from other trajectories in the condition, as in Eq.~\eqref{eq:11}. With this convention, define
\begin{equation}\label{eq:C:weighted}
C_{Y,\mathrm{late}}^{(w)}=
\frac{1}{|\mathcal C|}\sum_{c\in\mathcal C}
\frac{\sum_{i\in c}w_i\operatorname{Re}(u_{Y,i}u_{\mathrm{late},i}^{*})}
{\sum_{i\in c}w_i}.
\end{equation}
Here $Y\in\{\mathrm{obs},\mathrm{lin},\mathrm{exch}\}$ identifies the observed axis, the linear-reference axis, or the average over reconstructed axes; $\mathcal C$ contains all $27$ conditions or the nine conditions of one quench. The primary choice is $w_i=1$. The alternative weights are
\begin{equation}\label{eq:C:weights}
\begin{aligned}
w_i^{\mathrm{late}}&=|Q_{2,\mathrm{late},i}|,\\
w_i^{\mathrm{sym}}&=|Q_{2,\mathrm{late},i}|\sqrt{|Q_{2,\mathrm{obs},i}|\,|Q_{2,\mathrm{lin},i}|}.
\end{aligned}
\end{equation}
The weight $w_i$ remains attached to trajectory $i$ when errors are reassigned. Another trajectory $\ell$ supplies only the complex representation of its angular error, $\varepsilon_\ell=u_{\mathrm{lin},\ell}u_{\mathrm{obs},\ell}^{*}$. The uncorrected difference, $C_{\mathrm{obs},\mathrm{late}}^{(w)}-C_{\mathrm{lin},\mathrm{late}}^{(w)}$, compares the observed and linear-reference axes directly. The difference expected from reassigned errors, $C_{\mathrm{obs},\mathrm{late}}^{(w)}-C_{\mathrm{exch},\mathrm{late}}^{(w)}$, repeats the comparison after the measured reference error from trajectory $\ell$ has been applied to trajectory $i$ within the same condition. Subtracting this expected difference from the uncorrected difference gives the remainder in Eq.~\eqref{eq:residual_identity}. The observed term therefore cancels for every weight choice before any conditions are compared.

Table~\ref{tab:C:null} first reports how this comparison depends on the trajectory weight, both for the full ensemble and for the three quench groups. With equal trajectory weights, the remainder is near zero for the shallow quench, positive for the intermediate quench, and negative for the deep quench; each interval includes zero. Under symmetric trajectory weights, the actual linear-to-late alignment is $0.2638$, compared with $0.2263$ after error reassignment. Their difference gives the negative remainder $-0.0375$, with interval $[-0.0617,-0.0099]$: the actual linear reference is more closely aligned with the late axis than the reconstructed axis under this weighting. The interval that includes zero in the main text applies to the primary equal-trajectory comparison; symmetric weighting instead gives this negative remainder.

\begin{table}[H]
\centering\small
\caption{Error-reassignment comparison with common trajectory weights. The first two numerical columns report the uncorrected difference and the difference expected from reassigned errors; the final column gives the remainder and its paired $95\%$ bootstrap interval. Each row uses equal averaging over the physical conditions in that ensemble.}\label{tab:C:null}
\setlength{\tabcolsep}{4pt}
\begin{tabularx}{\linewidth}{@{}p{0.31\linewidth} r r >{\centering\arraybackslash}X@{}}
\toprule
Weight and ensemble & \shortstack{Observed $-$\\linear} & \shortstack{Expected from\\reassigned errors} & \shortstack{Remainder\\\textnormal{[95\% interval]}}\\
\midrule
Equal trajectory, full & $0.0575$ & $0.0626$ & \shortstack{$-0.0051$\\$[-0.0275,0.0178]$}\\
$r=-0.5$, equal trajectory & $0.0002$ & $0.0001$ & \shortstack{$0.0001$\\$[-0.0014,0.0018]$}\\
$r=-1$, equal trajectory & $0.0607$ & $0.0388$ & \shortstack{$0.0219$\\$[-0.0094,0.0515]$}\\
$r=-1.5$, equal trajectory & $0.1116$ & $0.1490$ & \shortstack{$-0.0374$\\$[-0.0968,0.0251]$}\\
Late amplitude, full & $0.0596$ & $0.0742$ & \shortstack{$-0.0146$\\$[-0.0392,0.0114]$}\\
Symmetric amplitude, full & $0.0493$ & $0.0868$ & \shortstack{$-0.0375$\\$[-0.0617,-0.0099]$}\\
\bottomrule
\end{tabularx}
\end{table}

\begin{figure}[H]
\centering
\includegraphics[width=0.95\linewidth]{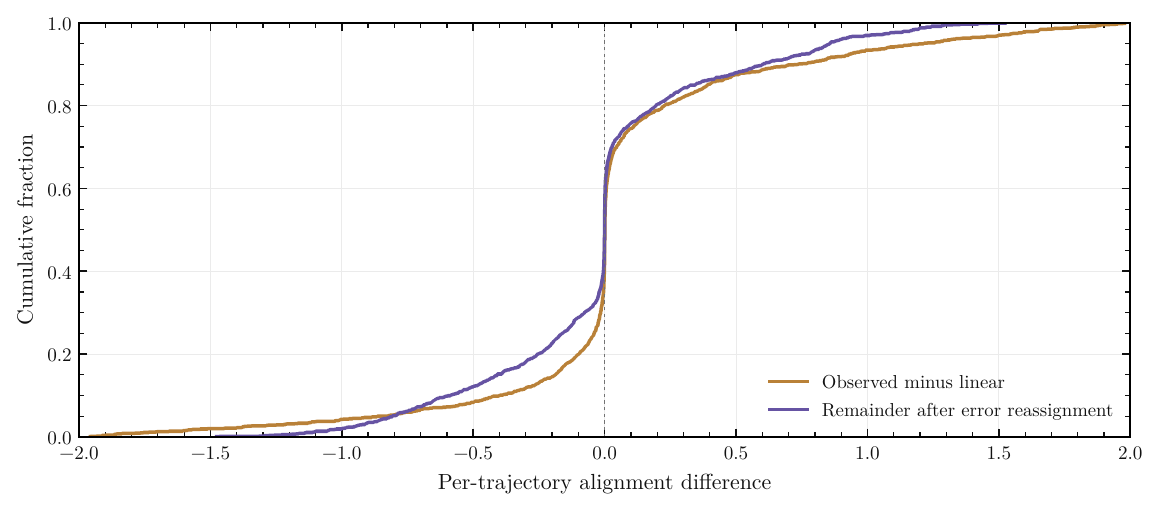}
\caption{Complete trajectory-level contribution distributions for the observed-minus-linear alignment difference and the difference remaining after error reassignment. Each empirical cumulative distribution contains all $1296$ trajectories with equal trajectory weights.}
\label{fig:C:contribution-cdf}
\end{figure}

To show how these averages arise from individual trajectories, Fig.~\ref{fig:C:contribution-cdf} gives the complete contribution distributions.

Intervals use $10000$ paired bootstrap samples. Within every condition, $48$ trajectory records are drawn with replacement, preserving each record's observed, linear, and late $Q_2$ values and its trajectory weight. The exact mean error over other sampled records is recomputed in each draw; the uncorrected difference and the difference expected from reassigned errors use the same draw. Percentile endpoints are the $2.5$th and $97.5$th percentiles. All spectral axes in this analysis are well defined because none of the complex $Q_2$ values vanishes: the minimum amplitudes of the observed, linear, and late values are $0.00215$, $0.00176$, and $0.00197$, respectively.

\subsection{Restricting which reference errors may be exchanged}\label{app:C:strata}

To test whether error reassignment is mixing trajectories that differ too strongly within the same physical condition, we restrict which trajectories may supply an error. Table~\ref{tab:C:strata} forms restricted pools using the observed amplitude, the magnitude $|\arg\varepsilon_i|$ of the doubled-angle error, or the early length fraction $L_1(30)/L_{\mathrm{box}}$. Median and quartile groups are formed by equal-count ranks in the original condition sample. A further check crosses the amplitude and error median groups. These original rank groups are fixed during bootstrap sampling; records are resampled within each group while its size is preserved, and expected alignments after reassignment are recomputed within the same group. The crossed grouping has at least four records in every group. The early length fraction spans $0.0739$--$0.3717$ across the original ensemble.

\begin{table}[H]
\centering\small
\caption{Sensitivity to restricting the source of reassigned errors for the full-ensemble remainder under equal trajectory weights. Every row preserves the physical condition and the original rank-group sizes. Intervals use $10000$ within-group bootstrap samples. The uncorrected point difference remains $0.0575$ in all rows.}\label{tab:C:strata}
\setlength{\tabcolsep}{5pt}
\begin{tabularx}{\linewidth}{@{}X >{\centering\arraybackslash}p{0.42\linewidth}@{}}
\toprule
Restriction within each condition & Remainder [95\% interval]\\
\midrule
Observed-amplitude median groups & $0.0014\; [-0.0205,0.0232]$\\
Observed-amplitude quartiles & $-0.0006\; [-0.0216,0.0199]$\\
Angular-error median groups & $-0.0033\; [-0.0216,0.0149]$\\
Angular-error quartiles & $0.0002\; [-0.0165,0.0166]$\\
Early $L_1(30)/L_{\mathrm{box}}$ median groups & $-0.0076\; [-0.0294,0.0147]$\\
Early $L_1(30)/L_{\mathrm{box}}$ quartiles & $-0.0039\; [-0.0250,0.0179]$\\
Amplitude $\times$ error median groups & $-0.0022\; [-0.0190,0.0160]$\\
\bottomrule
\end{tabularx}
\end{table}

All seven intervals span zero. These checks retain the measured error distribution within narrower classes of trajectories; they do not require a reflection-symmetric error distribution. As a separate randomization check, we generated $20000$ independent within-condition reassignments in which no trajectory retained its own error. They give a mean expected difference of $0.0627$, a central $95\%$ randomization range of $[0.0395,0.0858]$, and a one-sided probability $0.670$ of a randomized difference at least as large as the uncorrected value. The uncorrected difference is therefore within the range ordinarily produced by error reassignment alone. This randomization range describes variation among error reassignments, whereas the table intervals describe resampling the realization ensemble.

\subsection{Matching the remaining scale change}\label{app:C:matching}

To determine whether quench differences at a fixed time merely reflect different rates of coarsening, we compare trajectories with the same remaining change in length scale. The coordinate $s_c(t)$ in Eq.~\eqref{eq:8} compares each trajectory's current $L_1(t)$ with its mean late length and takes the median logarithmic ratio within the condition. Over $30\leq t<340$, every one of the $27$ condition curves covers the interval $[0.03077,0.05704]$. The three selected values $0.040$, $0.045$, and $0.050$ lie inside this intersection, so no condition is extrapolated when the quenches are compared. A condition curve can cross the same selected value more than once because $s_c(t)$ is not strictly monotonic. The primary rule uses the last crossing before $t=340$. It selects the later coarsening branch rather than an earlier transient, consistent with Sec.~\ref{sec:stage-comparison}, while keeping the comparison outside the late averaging window.

\begin{figure}[H]
\centering
\includegraphics[width=0.96\linewidth]{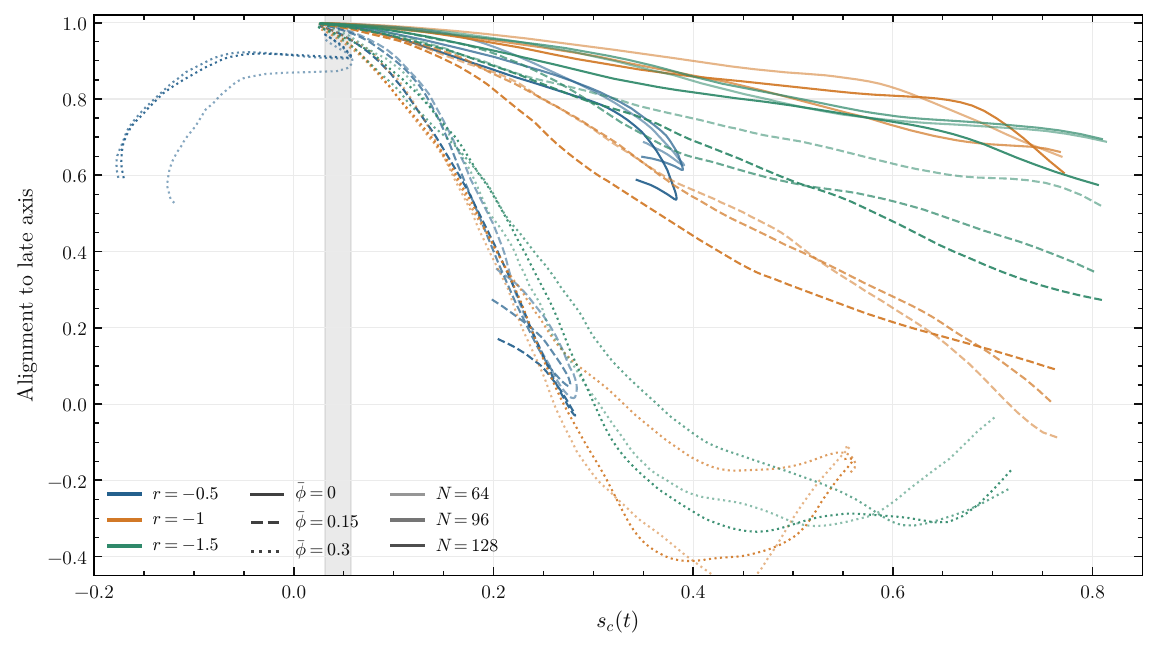}
\caption{All $27$ condition-level alignment curves against $s_c(t)$, drawn in time order. Color identifies quench depth, line style identifies mean composition, and lighter or darker shades distinguish box sizes. The gray band marks the interval covered by every condition curve and used for matching.}
\label{fig:C:condition-curves}
\end{figure}

We use two averages within each condition: the amplitude-product-weighted alignment of Eq.~\eqref{eq:7} and the equally weighted mean of the trajectory-level cosines. For either choice, we linearly interpolate the condition-level alignment between the two saved frames that bracket the selected crossing, using the same interpolation fraction as for the scale coordinate. We then average these matched alignments equally over conditions within each quench. Individual complex $Q_2$ values are not interpolated. Figure~\ref{fig:C:condition-curves} displays all condition curves and the interval shared by them.

\begin{table}[H]
\centering\small
\caption{Quench contrasts at fixed time and after matching at the last crossing of the selected scale coordinate before $t=340$. Each entry gives the contrast and its paired $95\%$ interval on separate lines. ``Product'' denotes amplitude-product weighting within conditions; ``equal'' denotes equal trajectory weights. S, I, and D denote shallow ($r=-0.5$), intermediate ($r=-1$), and deep ($r=-1.5$) quenches.}\label{tab:C:matched_contrasts}
\setlength{\tabcolsep}{3pt}
\begin{tabularx}{\linewidth}{@{}p{0.14\linewidth} p{0.12\linewidth} *{3}{>{\centering\arraybackslash}X}@{}}
\toprule
Comparison & Weight & I $-$ S & D $-$ S & D $-$ I\\
\midrule
$t=30$ & Product & \shortstack{$-0.3275$\\$[-0.4172,-0.2291]$} & \shortstack{$-0.1968$\\$[-0.2874,-0.1030]$} & \shortstack{$0.1307$\\$[0.0300,0.2319]$}\\
$s_c=0.040$ & Product & \shortstack{$0.0100$\\$[0.0037,0.0183]$} & \shortstack{$0.0119$\\$[0.0058,0.0199]$} & \shortstack{$0.0019$\\$[-0.0028,0.0067]$}\\
$s_c=0.045$ & Product & \shortstack{$0.0115$\\$[0.0039,0.0211]$} & \shortstack{$0.0139$\\$[0.0065,0.0233]$} & \shortstack{$0.0024$\\$[-0.0035,0.0084]$}\\
$s_c=0.050$ & Product & \shortstack{$0.0131$\\$[0.0041,0.0243]$} & \shortstack{$0.0160$\\$[0.0073,0.0269]$} & \shortstack{$0.0029$\\$[-0.0044,0.0102]$}\\
\addlinespace
$t=30$ & Equal & \shortstack{$-0.1863$\\$[-0.2706,-0.1006]$} & \shortstack{$-0.1166$\\$[-0.2019,-0.0290]$} & \shortstack{$0.0696$\\$[-0.0182,0.1605]$}\\
$s_c=0.040$ & Equal & \shortstack{$0.0541$\\$[0.0244,0.0850]$} & \shortstack{$0.0583$\\$[0.0297,0.0891]$} & \shortstack{$0.0042$\\$[-0.0151,0.0242]$}\\
$s_c=0.045$ & Equal & \shortstack{$0.0572$\\$[0.0256,0.0903]$} & \shortstack{$0.0618$\\$[0.0307,0.0942]$} & \shortstack{$0.0046$\\$[-0.0180,0.0267]$}\\
$s_c=0.050$ & Equal & \shortstack{$0.0606$\\$[0.0284,0.0965]$} & \shortstack{$0.0632$\\$[0.0308,0.0992]$} & \shortstack{$0.0027$\\$[-0.0221,0.0278]$}\\
\bottomrule
\end{tabularx}
\end{table}

Within the interval covered by all condition curves, both deeper quenches have higher alignment than the shallow quench, and the pointwise intervals for those contrasts exclude zero. Table~\ref{tab:C:matched_contrasts} also shows that the separation is larger with equal trajectory weights than with amplitude-product weights. This ordering is established only over the late-stage interval in which all $27$ curves overlap. To place the small quench differences beside the alignments being compared, Table~\ref{tab:C:matched_levels} reports the absolute levels at the middle selected value. For the primary last crossings, the mean matching times at $s_c=0.045$ are $320.1$, $318.0$, and $315.3$ for S, I, and D. Across all three selected values and conditions, matches occur at $304.2$--$329.2$, before the late window starts at $340$. The high absolute alignments accompany this proximity. These alignments are measured rather than imposed: $s_c=0$ means that the median logarithmic length ratio is zero; it does not require the axes to coincide.

\begin{table}[H]
\centering\small
\caption{Absolute alignment at $s_c=0.045$ and sensitivity to choosing the first or last crossing before $t=340$. Entries give point estimates and $95\%$ bootstrap intervals. S, I, and D have the same meaning as in Table~\ref{tab:C:matched_contrasts}. Only the three shallow-quench conditions with $\bar\phi=0.30$ change when the first crossing is used.}\label{tab:C:matched_levels}
\setlength{\tabcolsep}{3pt}
\begin{tabularx}{\linewidth}{@{}p{0.14\linewidth} p{0.12\linewidth} *{3}{>{\centering\arraybackslash}X}@{}}
\toprule
Crossing & Weight & S & I & D\\
\midrule
Last & Product & \shortstack{$0.9722$\\$[0.9632,0.9779]$} & \shortstack{$0.9837$\\$[0.9783,0.9872]$} & \shortstack{$0.9861$\\$[0.9814,0.9892]$}\\
Last & Equal & \shortstack{$0.8892$\\$[0.8602,0.9159]$} & \shortstack{$0.9464$\\$[0.9285,0.9615]$} & \shortstack{$0.9510$\\$[0.9346,0.9642]$}\\
First & Product & \shortstack{$0.9576$\\$[0.9439,0.9667]$} & \shortstack{$0.9837$\\$[0.9783,0.9872]$} & \shortstack{$0.9861$\\$[0.9814,0.9892]$}\\
First & Equal & \shortstack{$0.8669$\\$[0.8375,0.8951]$} & \shortstack{$0.9464$\\$[0.9285,0.9615]$} & \shortstack{$0.9510$\\$[0.9346,0.9642]$}\\
\bottomrule
\end{tabularx}
\end{table}

Only the three shallow-quench conditions with $(r,\bar\phi)=(-0.5,0.30)$, one at each box size, select a different part of the curve when the first rather than the last crossing is used. That choice increases the deeper-minus-shallow contrasts, but their intervals remain above zero under both weighting schemes. The choice of crossing therefore changes the size of the remaining separation without restoring the fixed-time ordering.

Matching intervals use $5000$ bootstrap draws of complete trajectory histories within each condition. Every draw recomputes the condition-median scale coordinate, the condition alignment, and the crossing. A quench estimate is retained only when every constituent condition has the required crossing; a paired contrast requires validity for both quenches in the same draw. At $s_c=0.050$, $4993$ of $5000$ draws satisfy this requirement for the shallow quench and contrasts involving it, for either crossing choice and either weight. The other quench estimates and the two smaller selected values retain all $5000$ draws. Draws without a required crossing are omitted from the relevant percentile interval; no extrapolated or filled-in crossing is used. All intervals in Tables~\ref{tab:C:matched_contrasts} and \ref{tab:C:matched_levels} are pointwise.

The matching coordinate is a retrospective measure of structural progress. Because it can be nonmonotonic and because the condition curves overlap only over a finite interval, the matched comparison applies only to the crossings observed within that shared interval.

\section{Prediction protocols and supporting controls}\label{app:D}

This appendix specifies how the prediction tests determine whether early measurements contain information about the late state of the same trajectory beyond that provided by the physical condition. It also records the controls used to identify whether any improvement comes from directional information, scalar morphology, model choice, or explicit condition information.

\subsection{Targets, partitions, and evaluation}\label{app:D:evaluation}

To measure trajectory-specific prediction rather than differences among the $27$ physical conditions, every model is compared with a baseline that already knows the physical condition. The scalar targets are the means of $L_1$, $A_S$, and $\rho_{\chi,-}$ over the $31$ stored times from $t=340$ to $400$; the complex target is $Q_{2,\mathrm{late}}$, obtained by averaging $Q_2$ over that window and therefore retaining both anisotropy magnitude and axis. The condition baseline for each scalar target is its training-condition median. For the complex target, the baseline is the point in the real--imaginary plane that minimizes the sum of distances to the training values in that condition. This point is the geometric median and is computed with a modified Weiszfeld iteration \cite{VardiZhang2000}. These baselines use the physical condition but no early measurements from the trajectory being predicted.

Table~\ref{tab:baselines} shows why the condition must be included before early measurements are tested. Relative to one global median, the condition-specific medians reduce MAE by $68.7\%$ for $L_1$, $31.1\%$ for $A_S$, and $81.9\%$ for $\rho_{\chi,-}$. The remaining prediction tests therefore ask whether an early observation adds information beyond quench depth, mean composition, and box size.

\begin{table}[H]
\centering\small
\caption{\textbf{Baseline prediction from physical conditions.} Training-only condition medians are compared with training-only global medians. Errors retain each target's native units.}
\label{tab:baselines}
\begin{tabular}{@{}lrrr@{}}\toprule
\shortstack[l]{Late scalar\\target} & \shortstack[r]{Global-median\\MAE} & \shortstack[r]{Condition-median\\MAE} & \shortstack[r]{Reduction from\\condition knowledge}\\\midrule
$L_1$ & $1.47$ & $0.461$ & $68.7\%$\\
$A_S$ & $0.0802$ & $0.0553$ & $31.1\%$\\
$\rho_{\chi,-}$ & $1.01\times10^{-3}$ & $1.83\times10^{-4}$ & $81.9\%$\\
\bottomrule\end{tabular}
\end{table}

To evaluate trajectories that were kept out of fitting, we divide the data into five parts, use four parts for fitting, and use the remaining part for testing. This five-part division is repeated five times, with trajectories from every physical condition present in every part. Each trajectory therefore receives five predictions made without using that trajectory for fitting. Model predictions are averaged across repetitions before taking the absolute error; the corresponding five condition-baseline predictions are averaged in the same way. Equation~\eqref{eq:mae-skill-score} then defines the target-specific MAE-based skill score $\mathrm{SS}_{\mathrm{MAE},j}$ from these errors. For the three scalar targets, we report their equally weighted mean score, $\overline{\mathrm{SS}}_{\mathrm{MAE}}=\frac13\sum_{j=1}^{3}\mathrm{SS}_{\mathrm{MAE},j}$. Averaging the three target-specific scores prevents quantities with different physical units from being combined in one unscaled error.

All condition medians or geometric medians, input transformations, response scales, and normalization parameters are estimated from the four fitting parts. When a model setting must be selected, those fitting data are divided again into four parts, so the test trajectories remain unused. The $50$ scalar inputs comprise the ten quantities in Table~\ref{tab:A:features} at $t=0,10,20,30$, followed by their ten changes from $t=0$ to $t=30$, as specified in Appendix~\ref{app:A:aggregation}. The complex history contains real and imaginary parts of $Q_2$ at $t=0,6,14,30$, giving eight real inputs. The single-time complex input is $Q_2(30)$, whereas the amplitude-only input is $|Q_2(30)|$. The other-trajectory-direction input keeps the predicted trajectory's value of $|Q_2(30)|$ but combines it with the direction from another trajectory in the same physical condition and test part. All three comparisons use the same complex target and condition baseline.

Uncertainty intervals use paired bootstrap sampling, which repeatedly draws trajectories with replacement within each condition \cite{Efron1979}. We use $5000$ such samples for complex-target comparisons and $10000$ for scalar comparisons. Each sample keeps the observed target, model prediction, and baseline prediction from one trajectory together. The $95\%$ limits are the $2.5$th and $97.5$th percentiles of the resulting values. The procedure resamples the stored averaged predictions for test trajectories; model fitting and setting selection are not repeated. Paired differences in the skill score use the same bootstrap sample for both estimators. The ranges across repetitions or assignments of other-trajectory directions reported below are sensitivity summaries rather than confidence intervals.

\subsection{Model fitting and capacity choices}\label{app:D:models}

To check that a predictive conclusion does not depend on one model form or an arbitrary complexity choice, the scalar predictions use three model families. Ridge is a linear regression with a penalty on large coefficients \cite{HoerlKennard1970}. Extra Trees averages many regression trees whose splits are partly randomized \cite{Geurts2006}. Histogram gradient boosting builds a sequence of regression trees from binned inputs, with each tree correcting the preceding prediction. Each model fits deviations from the training-condition target medians. During fitting, each target residual is divided by its training standard deviation, and this transformation is inverted when the condition baseline is added back. Ridge also standardizes the inputs within the training partition, whereas the two tree-based models use the raw input columns. The four-part division of the fitting data selects one complexity setting by minimizing the equally weighted mean, over the three targets, of the model MAE divided by the corresponding condition-baseline MAE. This criterion matches the target-wise comparison reported above while retaining each model's stated fitting objective.

The scalar Ridge grid is $\alpha=10^{-4+0.5h}$ for $h=0,\ldots,16$. The two tree-based model families each test the three complete settings in Table~\ref{tab:D:capacity}. Each row is one candidate, with its parameters varied together. Extra Trees uses $128$ trees, all input features eligible at each split, no bootstrap subsampling of training rows, and a squared-error splitting criterion. Histogram gradient boosting fits one model per target with absolute-error loss, learning rate $0.05$, and early stopping disabled; the capacity row is selected jointly across targets. Ridge and Extra Trees retain squared-error training objectives even though selection and evaluation use MAE.

\begin{table}[H]
\centering\small
\caption{The complete scalar tree-model candidate sets. Each row is one configuration, not part of a Cartesian product. ``Unlimited'' means no specified depth limit. Extra Trees has $128$ trees in every row. For boosting, the iteration count is the number of boosting stages and the $L_2$ entry is the leaf-value regularization strength.}\label{tab:D:capacity}
\begin{tabularx}{\linewidth}{@{}Xrr@{}}
\toprule
Extra Trees setting & Maximum depth & Minimum leaf size\\
\midrule
Low capacity & $4$ & $16$\\
Intermediate capacity & $8$ & $8$\\
High capacity & Unlimited & $5$\\
\bottomrule
\end{tabularx}
\medskip
\begin{tabularx}{\linewidth}{@{}Xrrrrr@{}}
\toprule
Boosting setting & Iterations & \shortstack{Maximum\\leaves} & \shortstack{Maximum\\depth} & \shortstack{Minimum\\leaf size} & $L_2$\\
\midrule
Low capacity & $120$ & $7$ & $3$ & $24$ & $1.0$\\
Intermediate capacity & $200$ & $15$ & $5$ & $16$ & $1.0$\\
High capacity & $260$ & $31$ & Unlimited & $10$ & $0.1$\\
\bottomrule
\end{tabularx}
\end{table}

For the complex target, Ridge standardizes its inputs inside the fitting data and predicts the two-coordinate late residual from the training-condition geometric median. Dividing those data into four parts selects $\alpha=10^{-4+h}$, $h=0,\ldots,8$, by MAE in the complex plane. The corresponding Extra Trees model instead uses fixed settings: $128$ trees, maximum depth $4$, minimum leaf size $16$, all features eligible, and squared-error fitting. The single-time, amplitude-only, and other-trajectory-direction inputs retain the fitting procedure of their model family.

\subsection{Condition information and prediction averaging}\label{app:D:scalar-controls}

To test whether the scalar models appear weak only because their morphology-based correction is shared across conditions, we also allow that correction to depend explicitly on condition. In the primary scalar models, the exact condition sets the baseline median, while one model shared across conditions uses morphology features to predict the remaining deviation. For this control, Ridge is fitted separately inside each condition, with separate training-only input standardization and a single regularization strength selected using the four-part setting-selection criterion. The two tree learners append a $27$-component indicator vector, with one active entry identifying the condition, to the $50$ morphology features, giving $77$ inputs. The late targets, repeated training/test partitions, baselines, and order of averaging predictions before evaluating loss remain unchanged.

\begin{table}[H]
\centering\small
\caption{Scalar prediction when the morphology-based correction also uses the exact condition. Values are equally weighted mean MAE-based skill scores in percent relative to the training-condition median, with paired $95\%$ intervals. The three intervals include zero.}\label{tab:D:condition-residual}
\begin{tabularx}{\linewidth}{@{}Xrr@{}}
\toprule
Morphology-based correction & $\overline{\mathrm{SS}}_{\mathrm{MAE}}$ (\%) & $95\%$ interval (\%)\\
\midrule
Condition-specific Ridge & $0.239$ & $[-0.361,\ 0.831]$\\
Extra Trees with 27-component condition indicator & $0.364$ & $[-0.189,\ 0.906]$\\
Absolute-error boosting with 27-component condition indicator & $-0.132$ & $[-0.607,\ 0.339]$\\
\bottomrule
\end{tabularx}
\end{table}

This comparison supports the interpretation that the small amount of additional information in the scalar histories does not arise simply from omitting the condition from the morphology-based correction. In every model comparison throughout the manuscript, we first average the five test predictions and their corresponding baselines and then evaluate the absolute error and skill score.

\subsection{Direction controls, history, and complex nonnegative-slope models}\label{app:D:complex-controls}

To isolate the information carried by the trajectory's own early direction, this control preserves $|Q_2(30)|$ but replaces its direction with that of another trajectory. The resulting input has two real components, the same number as the full complex value $Q_2(30)$.

\begin{figure}[H]
\centering
\includegraphics[width=\linewidth]{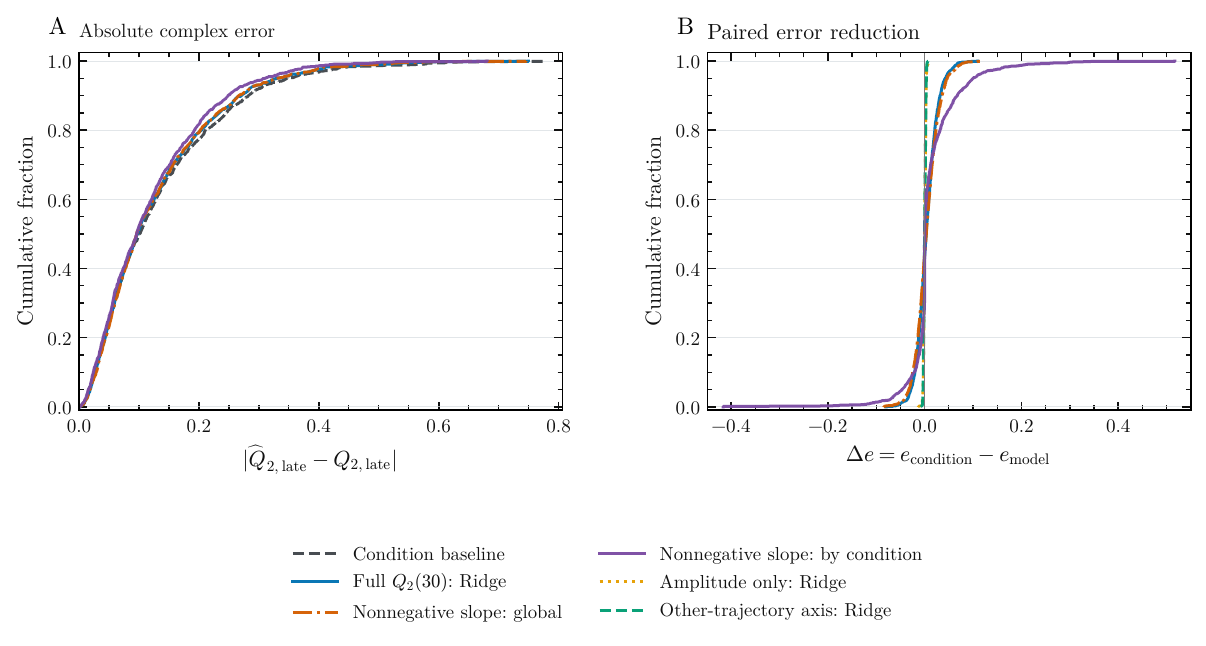}
\caption{Complex-prediction error distributions. (A) Absolute errors for the condition baseline, the single-time Ridge model, and the global and condition-specific nonnegative-slope models. (B) Paired error reductions relative to the condition baseline, including the amplitude-only input and the first prespecified scheme for replacing directions. Each curve contains all $1296$ trajectory contributions.}
\label{fig:D:complex-cdfs}
\end{figure}

We evaluate five prespecified replacement schemes. Within each scheme, a separate direction assignment is constructed for each of the five repetitions of the training/test division. Each assignment pairs a trajectory with a different trajectory in the same physical condition and in the same part of that repetition's division, without using late-target information. For each scheme, the five repeated predictions are averaged before MAE is evaluated. Predictions from different schemes are not averaged together. Figure~\ref{fig:D:complex-cdfs} shows the first prespecified scheme, and Table~\ref{tab:D:donor-range} reports the range across all five.

The maximum amplitude-preservation error across these replacements was $1.11\times10^{-16}$. Table~\ref{tab:D:donor-range} gives the sensitivity to the selected replacement directions for both the Ridge and Extra Trees models and the nonnegative-slope benchmarks.

\begin{table}[H]
\centering\small
\caption{MAE-based skill score ranges across five prespecified schemes for replacing each trajectory's direction with that of another trajectory in the single-time complex input. Each scheme contains one assignment per training/test repetition and preserves the same target, condition baseline, and fitting protocol. Ranges describe sensitivity to the replacements and are not confidence intervals.}\label{tab:D:donor-range}
\begin{tabularx}{\linewidth}{@{}Xr@{}}
\toprule
Predictor supplied with other-trajectory directions & $\mathrm{SS}_{\mathrm{MAE},j}$ range (\%)\\
\midrule
Ridge & $[-0.0515,\ 0.0221]$\\
Extra Trees & $[-0.0618,\ 0.0597]$\\
Global nonnegative-slope model & $[-0.0079,\ 0.0087]$\\
Condition-specific nonnegative-slope models & $[-0.3145,\ 0.0407]$\\
\bottomrule
\end{tabularx}
\end{table}

For a directly interpretable test of directional persistence, the model in Eq.~\eqref{eq:complex-persistence-map}, $\widehat Q_{2,\mathrm{late}}=b_c+\lambda Q_2(30)$, adds the early complex $Q_2$ value to the condition baseline with either one global nonnegative real slope or one slope per condition. The slopes minimize the sum of absolute distances in the complex plane over the fitting data, with no extra intercept or fitted rotation. The late response is centered by $b_c$, whereas the early complex input remains uncentered. A zero slope recovers the same condition baseline. A positive slope adds a correction along the observed early direction and therefore tests whether that correction reduces late-target error. The versions using other-trajectory directions retain the same fitting procedure and parameter counts.

The fitted global slope ranges from $0.313$ to $0.369$ across the $25$ training fits. The $675$ condition-by-part slopes range from $0$ to $1.684$, with $169$ at the zero boundary. Values $\lambda_c>1$ amplify the directional correction relative to the condition baseline: $|\widehat Q_{2,\mathrm{late}}-b_c|=\lambda_c|Q_2(30)|$. Allowing the slope to depend on condition therefore permits the size of the directional correction to differ among conditions instead of imposing one common slope.

Table~\ref{tab:D:paired-complex} compares models by paired changes in the MAE-based skill score, complementing the absolute values in the main text. Adding the three earlier complex observations gives no improvement over $Q_2(30)$ in either tested family; the Ridge difference is below zero. The intervals do not show either the single-time Ridge model or the single-time Extra Trees model outperforming the global nonnegative-slope model. For the input sets tested here, these paired results show that the latest early complex $Q_2$ contains the useful predictive information.

\begin{table}[H]
\centering\small
\caption{Paired complex-prediction comparisons. A positive difference favors the first estimator named. Differences and $95\%$ intervals are in percentage points of the complex-MAE skill score and use $5000$ paired bootstrap replicates. The four-time history is $t=0,6,14,30$.}\label{tab:D:paired-complex}
\begin{tabularx}{\linewidth}{@{}Xrr@{}}
\toprule
Comparison & Difference (pp) & $95\%$ interval (pp)\\
\midrule
Ridge: history minus single time & $-0.628$ & $[-1.093,\ -0.176]$\\
Extra Trees: history minus single time & $-0.273$ & $[-0.616,\ 0.063]$\\
Single-time Ridge minus global nonnegative-slope model & $-0.110$ & $[-0.271,\ 0.047]$\\
Single-time Extra Trees minus global nonnegative-slope model & $-0.316$ & $[-0.665,\ 0.018]$\\
\bottomrule
\end{tabularx}
\end{table}

\subsection{Target-specific scalar nonnegative-slope models}\label{app:D:scalar-maps}

To test whether a scalar target's own early value contains corresponding information about its late value, we use that target's value at $t=30$ as the only trajectory-specific input. For an early scalar $x$ and its late-window mean $y$, the fitted nonnegative-slope model is
\begin{equation}\label{eq:D:scalar-map}
\widehat y=m_{y,c}+\beta\,[x(30)-m_{x,c}],\qquad \beta\geq0,
\end{equation}
where both medians are estimated from the fitting data. One version fits a global slope for each target; another fits one slope per target and condition. Slopes minimize scalar MAE over the fitting data, without a separate setting-selection step. Zero slope exactly recovers the target's condition-median baseline. Unlike the complex-target model above, this construction centers both the early predictor and the late response.

The same five repeated five-part divisions and $10000$ paired bootstrap samples give Table~\ref{tab:D:scalar-maps}. These nonnegative-slope models are exploratory controls that test whether an early value of each target reduces the error of its own late value. The table retains all three scalar outcomes so that the small gain for $A_S$ can be assessed alongside the scale and topology results.

\begin{table}[H]
\centering\small
\caption{Target-specific scalar nonnegative-slope models. MAE-based skill scores and $95\%$ intervals are in percent relative to the training-condition median. Boundary counts give zero-slope fits out of $25$ global fits or $675$ condition-by-part fits for each target.}\label{tab:D:scalar-maps}
\begin{tabularx}{\linewidth}{@{}Xlrrr@{}}
\toprule
Slope specification & Target & $\mathrm{SS}_{\mathrm{MAE},j}$ (\%) & $95\%$ interval (\%) & \shortstack{Zero-slope\\fits}\\
\midrule
Global & $L_1$ & $0.009$ & $[-0.109,\ 0.124]$ & $1/25$\\
Global & $A_S$ & $0.683$ & $[0.053,\ 1.293]$ & $0/25$\\
Global & $\rho_{\chi,-}$ & $0.347$ & $[-0.171,\ 0.873]$ & $0/25$\\
\addlinespace
Condition-specific & $L_1$ & $-0.311$ & $[-1.050,\ 0.394]$ & $318/675$\\
Condition-specific & $A_S$ & $0.492$ & $[-1.631,\ 2.516]$ & $205/675$\\
Condition-specific & $\rho_{\chi,-}$ & $0.409$ & $[-1.333,\ 2.046]$ & $354/675$\\
\bottomrule
\end{tabularx}
\end{table}

\begin{figure}[H]
\centering
\includegraphics[width=\linewidth]{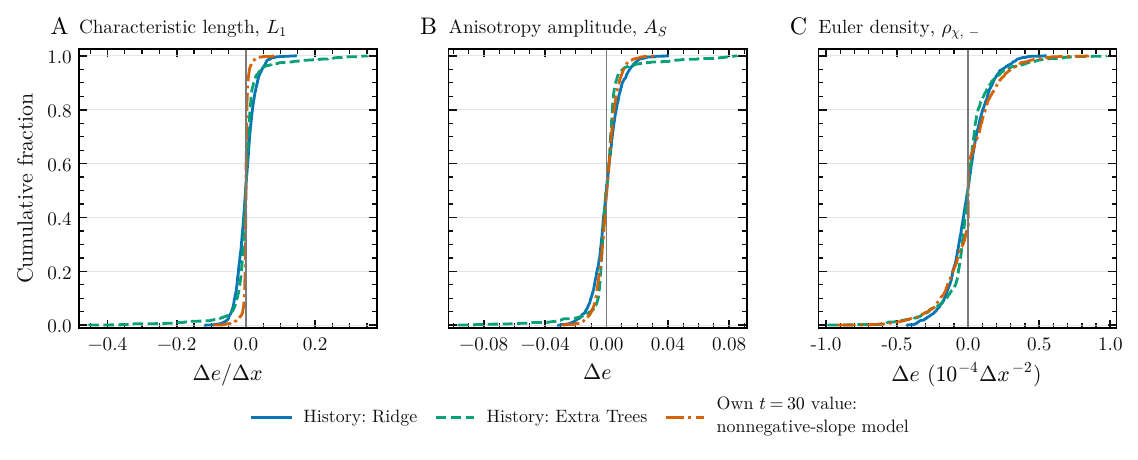}
\caption{Paired error-reduction distributions for the late scalar targets $L_1$, $A_S$, and $\rho_{\chi,-}$. Each panel compares the Ridge and Extra Trees models using scalar histories with the target-specific global nonnegative-slope model; positive values indicate lower error than the training-condition median.}
\label{fig:D:scalar-distributions}
\end{figure}

The global scalar models give a mean targetwise MAE reduction of $0.347\%$ relative to the condition baselines, with interval $[0.065,0.623]\%$; the condition-specific version gives $0.197\%$ with interval $[-0.804,1.126]\%$. The difference in equally weighted mean score between the global model and Ridge is $0.288$ percentage points, with interval $[-0.049,0.611]$. The corresponding difference between the global model and Extra Trees is $0.064$ percentage points, with interval $[-0.410,0.544]$. The global $A_S$ slope is positive in every training fit, ranging from $0.089$ to $0.171$ with median $0.147$. Thus the target-specific model reveals a weak association between early and late anisotropy magnitude, whereas the intervals for separate slopes by condition include zero for every scalar target. Figure~\ref{fig:D:scalar-distributions} shows the corresponding trajectory-level error-reduction distributions.

\bibliographystyle{unsrtnat}
\bibliography{references}

\end{document}